\documentclass[a4paper,12pt]{article}
\usepackage{amsmath, amsthm}
\usepackage{amssymb,amsfonts}
\usepackage{amsmath,amssymb}
\usepackage[mathscr]{euscript}
\usepackage{bm}
\usepackage{enumerate}
\usepackage{booktabs}
\usepackage{csquotes}
\usepackage{fancyhdr}
\usepackage{graphicx}
\usepackage{hyperref}
\usepackage{authblk}

\usepackage{amssymb}

\theoremstyle{definition}
\newtheorem{definition}{Definition}[section]

\theoremstyle{plain}
\newtheorem{theorem}[definition]{Theorem}
\newtheorem{lemma}[definition]{Lemma}
\newtheorem{proposition}[definition]{Proposition}

\numberwithin{equation}{section}

\usepackage{color}   
\usepackage{hyperref}
\usepackage{url}
\begin{document}
	\date{}
	\author[1]{Wajid M. Shaikh}
	\author[2]{Rupali S. Jain}
	\author[3]{B. Surendranath Reddy}
	\author[4]{Bhagyashri S. Patil}
	\affil[1]{Research Scholar, School of Mathematical Sciences, SRTMU Nanded}
	\affil[2]{Associate Professor, School of Mathematical Sciences, SRTMU Nanded}
	\affil[3]{Assistant Professor, School of Mathematical Sciences, SRTMU Nanded}
	\affil[4]{Research Scholar, School of Mathematical Sciences, SRTMU Nanded}
	\title{
		{Construction of Minimal Binary Linear Codes of dimension $n+3$}\\
	}
	\maketitle
	\begin{abstract}
		
		In this paper, we will give the generic construction of a binary linear code of dimension $n+3$ and derive the necessary and sufficient conditions for the constructed code to be minimal. Using generic construction, a new family of minimal binary linear code will be constructed from a special class of Boolean functions violating the Ashikhmin-Barg condition. We also obtain the weight distribution of the constructed minimal binary linear code.
	\end{abstract}
	\section{Introduction}

Linear codes serve as a cornerstone of modern communication systems, providing robust error control mechanisms essential for reliable data transmission in various applications ranging from telecommunications to digital broadcasting. Minimal linear codes is a special class of linear codes widely used in secret sharing, data storage, and information processing systems in modern technology \cite{1}\cite{2}. So, the construction of minimal linear codes is one of the thrust area for many researchers in coding theory.
	\par Minimal codes are characterized by the property that none of the codeword is covered by some other linearly independent codeword. There are two approaches to studying minimal linear codes one is the algebraic method and another is geometric method. The algebraic method is based on the Hamming weights of the codeword. In 1998,  Ashikhmin and Barg\cite{11} gave a sufficient condition for linear code to be minimal, which is stated below.
	\begin{lemma}\cite{11}\label{lemma1.1}[Ashikhmin-Barg]
	Let $p$ be a prime. Then a linear code $\mathscr{C}$ over $\mathbb{F}_{p}$ is minimal if
	\begin{equation*}
	\frac{{wt}_{min}}{{wt}_{max}}>\frac{p-1}{p}
	\end{equation*}
where, ${wt}_{min}$ and ${wt}_{max}$ denote the minimum and maximum non-zero Hamming weights for $\mathscr{C}$, respectively.
	\end{lemma}
\par From the above lemma, many minimal linear codes constructed but constructing  minimal linear codes violating Ashikhmin-Barg condition is an interesting research topic. In the year 2018, Chang and Hyun\cite{12} constructed an infinite family of minimal binary linear codes, which are not satisfying Ashikhmin-Barg condition with the help of mathematical objects called simplicial complexes and generic construction
\begin{equation}\label{eq1.1}
\mathscr{C}_{f}=\{(\alpha f({\bf x})+{\bf w}\cdot {\bf x})_{\bf x\in \mathbb{F}_{p}^{n*}}\ | \ \alpha\in \mathbb{F}_{p}, \ {\bf w}\in \mathbb{F}_{p}^{n}\}
\end{equation}
The necessary and sufficient condition for binary linear code to be minimal was given by Ding, Heng, and Zhou \cite{3} in 2018, and using special Boolean functions, they constructed three classes of minimal binary linear codes. Heng et al. \cite{4} discovered a necessary and sufficient condition for linear codes over finite fields to be minimum. This is a generalised version of result Ding, C., Heng, Z., \& Zhou, Z. \cite{3}. Also, several researchers constructed minimal linear code from different kinds of functions (see \cite{6},\cite{7},\cite{8},\cite{9}, and \cite{10}).
\par Till date, several minimum linear codes have been constructed that do not meet the Ashikhmin-Barg condition, but the dimension of these codes is either $n$ or $n+1$, which reduces the information rate of a code. Recently, Liu and Liao \cite{13} constructed a minimal binary linear code using a Boolean function with dimension $n+2$.
\par \par In this paper, we extend the results of \cite{13} to dimension $n+3$. This paper is organised as follows: In Section 2, we give some basic definitions and results, which are prerequisites. Section 3 contains the generic construction of minimal binary linear code with dimension $n+3$ and give a necessary and sufficient conditions for this code to be minimal. In Section 4, we obtain a family of minimal linear codes and one special class of minimal linear codes, which violets Ashikhmin-Barg condition.
\section{Preliminaries}
In this section, we give some basic definitions and results that are prerequisites to this paper. We denote a finite field with $p$ elements by $\mathbb{F}_{p}$. A $k$-dimensional subspace of $\mathbb{F}_{p}^{n}$ is a $[n,k,d]$ linear code denoted by $\mathscr{C}$, where $d$ is the minimum Hamming distance. Let $A_{i}$ denote the count of the codeword having weight $i$ in $\mathscr{C}$ of length $n$. The weight enumerator of a code $\mathscr{C}$ is defined as $1+A_{1}z+A_{2}z+\cdots + A_{n}z^{n}$, and the sequence $(1,A_{1},A_{2},\ldots, A_{n})$ is called the weight distribution of a code $\mathscr{C}$. If the number of non-zero $A_{i}$'s is $t$, then $\mathscr{C}$ is called the $t$-weight code.
\begin{definition}[$Support$]\cite{14}
The $support$ of a codeword ${\bf c}$ is denoted by $\text{Supp}({\bf c})$, is the set of the coordinates at which ${c}$  is non-zero. i.e.
\begin{equation*}
	\text{Supp}({\bf c})=\{1\leq i\leq n \ | \ c_{i}\neq 0\}
\end{equation*}
\end{definition}
Note that, $\text{wt}({\bf c})=|\text{Supp}({\bf c})|$, where $\text{wt}({\bf x})$ is the Hamming weight of ${\bf c}$. For ${\bf y}\in \mathbb{F}_{p}^{n}$, if $\text{Supp}({\bf y})\subseteq \text{Supp}({\bf x})$, then we call ${\bf x}$ covers ${\bf y}$ and denoted by ${\bf y}\preceq {\bf x}$.
\begin{definition}
A codeword ${\bf x}$ is minimal if ${\bf x}$ covers only the codeword of the form $\alpha {\bf x}$, where $\alpha\in \mathbb{F}_{p}^{*}$. A code $\mathscr{C}$ is called minimal if every codeword in $\mathscr{C}$ is minimal.
\end{definition}
A Boolean function $f$ is a function from $\mathbb{F}_{2}^{n}$ to $\mathbb{F}_{2}$. The Walsh transform for a Boolean function is defined as follows
\begin{equation*}
	\widehat{f}({\bf w})=\sum_{{\bf x}\in\mathbb{F}_{2}^{n}}(-1)^{f({\bf x})+{\bf w}\cdot{\bf x}}
\end{equation*}
where, ${\bf w} \in \mathbb{F}_{2}^{n}$ and ${\bf w}\cdot{\bf x}$ is the standard inner product of ${\bf w}$ and ${\bf x}$. Another related walsh transform of $f$ is given by
\begin{equation*}
	\tilde{f}({\bf w})=\sum_{{\bf x}\in\mathbb{F}_{2}^{n}}f({\bf x})(-1)^{{\bf w}\cdot{\bf x}}
\end{equation*}
 where, $f({\bf x})$ is a real valued function which takes only values $0$ and $1$. Following lemma describe the relation between above defined Walsh transforms.
\begin{lemma}\label{lemma2.1}\cite{3}
Let $f: \mathbb{F}^{n}_{2}\rightarrow \mathbb{F}_{2}$ be a Boolean function. Then
\begin{equation*}
	\widehat{f}({\bf w})=\begin{cases}
		2^{n}-2\tilde{f}({\bf 0}),& \text{ if } {\bf w}={\bf 0}\\
		-2\tilde{f}({\bf w}), & \text{ if }, {\bf w}\neq {\bf 0}
	\end{cases}
\end{equation*}
\end{lemma}
\begin{lemma}\label{lemma2.2}\cite{3}
Let ${\bf x},{\bf y}\in \mathbb{F}_{2}^{n}$. Then ${\bf y}\preceq {\bf x}$ if and only if
\begin{equation*}
	\text{wt}({\bf x}+{\bf y})=\text{wt}({\bf x})-\text{wt}({\bf y})
\end{equation*}
\end{lemma}
	\section{Construction of Minimal Binary Linear Codes of dimension $n+3$ from a class of Boolean Functions}
	In this section, we extend and generalise the construction of minimal binary linear codes from the class of Boolean functions. Let $f,g$ and $h$ be Boolean functions with domain $\mathbb{F}_{2}^{n}$ and denote the collection of the functions and their linear combinations as $\mathcal{F}=\{\alpha f+\beta g+\gamma h \ | \ \alpha,\beta,\gamma \in \mathbb{F}_{2}\}\setminus\{0\}$ such that the following conditions hold:
	\begin{enumerate}
		\item $\phi$ is non-zero function, for all $\phi \in \mathcal{F}$.
		\item $\phi({\bf 0})=0$, for all $\phi\in \mathcal{F}$.
		\item $\phi_{1}\neq \phi_{2}$, for all $\phi_{1},\phi_{2}\in \mathcal{F}$.
	\end{enumerate}
	In this paper, we define a linear code $\mathscr{C}_{f,g,h}$, which is a generalization of the linear code defined in \cite{13} as follows:
	\begin{equation}\label{eq31}
		\mathscr{C}_{f,g,h}=\{c_{f,g,h}({\bf w})=(\alpha f({\bf x})+\beta g({\bf x})+\gamma h({\bf x})+{\bf w}\cdot {\bf x} )_{{\bf x}\in \mathbb{F}_{2}^{n*}} \ | \ \alpha,\beta,\gamma\in \mathbb{F}_{2}, \ {\bf w}\in \mathbb{F}_{2}^{n}\}
	\end{equation}
	\begin{theorem}\label{thm31}
		
		Let $\mathscr{C}_{f,g,h}$ be a binary code as defined above. Then, if $\phi({\bf x})\neq {\bf v}\cdot {\bf x} \ \forall \ {\bf v}\in \mathbb{F}_{2}^{n}$ and $\forall \ \phi\in \mathcal{F}$, then $\mathscr{C}_{f,g,h}$ has length $2^{n}-1$ and dimension $n+3$. Moreover, the weight distribution of $\mathscr{C}_{f,g,h}$ is given by the following multiset union:
		$\displaystyle\bigcup_{\phi \in \mathcal{F}}\left\{\frac{2^{n}-\widehat{\phi}({\bf w})}{2}| \ {\bf w}\in\mathbb{F}_{2}^{n} \right\}\bigcup \{2^{n-1}\ | \ {\bf w}\in \mathbb{F}_{2}^{n}\}\bigcup \{0\}$.
		\begin{proof}
			Clearly, the length of the code $\mathscr{C}_{f,g,h}$ is $2^{n}-1$. Let $\{{\bf u}_{1},{\bf u}_{2},\ldots, {\bf u}_{n}\}$ be a basis of $\mathbb{F}_{2}^{n}$. Then, we prove that $\{c_{f,g,h}({\bf u_{1}}),c_{f,g,h}({\bf u_{2}}),\ldots, c_{f,g,h}({\bf u_{n}}),c_{f}({\bf 0}),c_{g}({\bf 0}),c_{h}({\bf 0})\}$ is the largest linear independent set of $\mathscr{C}_{f,g,h}$. Consider,
			\begin{align*}
				\sum_{i=1}^{n}\delta_{i}c_{f,g,h}({\bf u_i})+\delta_{n+1}c_{f}({\bf 0})+\delta_{n+2}c_{g}({\bf 0})+\delta_{n+3}c_{h}({\bf 0})&={\bf 0}\\
				\sum_{i=1}^{n}\delta_{i}(f({\bf x})+g({\bf x})+h({\bf x})+{\bf u_i}\cdot{\bf x})_{{\bf x}\in \mathbb{F}_{2}^{n}}+\delta_{n+1}(f({\bf x}))_{{\bf x}\in \mathbb{F}_{2}^{n}}\notag\\+\delta_{n+2}(g({\bf x}))_{{\bf x}\in \mathbb{F}_{2}^{n}}+\delta_{n+3}(h({\bf x}))_{{\bf x}\in \mathbb{F}_{2}^{n}}&={\bf 0}\\
				\sum_{i=1}^{n}\delta_{i}({\bf u_i}\cdot{\bf x})_{{\bf x}\in \mathbb{F}_{2}^{n}}+(\delta_{1}+\delta_{2}+\cdots +\delta_{n}+ \delta_{n+1})(f({\bf x}))_{{\bf x}\in \mathbb{F}_{2}^{n}}+\notag\\(\delta_{1}+\delta_{2}+\cdots +\delta_{n}+\delta_{n+2})(g({\bf x}))_{{\bf x}\in \mathbb{F}_{2}^{n}}+(\delta_{1}+\delta_{2}+\cdots +\delta_{n}+\delta_{n+3})(h({\bf x}))_{{\bf x}\in \mathbb{F}_{2}^{n}}&={\bf 0}\\
				\sum_{i=1}^{n}\delta_{i}({\bf u_i}\cdot{\bf x})_{{\bf x}\in \mathbb{F}_{2}^{n}}+\alpha_{1}f({\bf x})+\alpha_{2}g({\bf x})+\alpha_{3}h({\bf x})&=0
			\end{align*}
		If $\alpha_{i}=0$, for $i\in \{1,2,3\}$, then $\delta_{j}=0$, for all $j$. If $\alpha_{i}\neq 0$, for some $i$, then $\phi({\bf x})={\bf v}\cdot{\bf x}$, for some $\phi\in \mathcal{F}$.\\
			From this, we get $\{c_{f,g,h}({\bf u_{1}}),c_{f,g,h}({\bf u_{2}}),\ldots, c_{f,g,h}({\bf u_{n}}),c_{f}({\bf 0}),c_{g}({\bf 0}),c_{h}({\bf 0})\}$ is linearly independent. From the dimension of $\mathbb{F}_{2}^{n}$, we can prove that any subset of $\mathscr{C}_{f,g,h}$ containing $n+4$ elements is linearly dependent.\\
			 Now we will give the weight enumeration  for the case $\alpha=\beta=\gamma=1$, we have\\
			\begin{align*}
				\widehat{f+g+h}({\bf w})&=\sum_{{\bf x}\in\mathbb{F}_{2}^{n}}(-1)^{f({\bf x})+g({\bf x})+h({\bf x})+{\bf w}\cdot{\bf x}}\\
				&=2^{n}-2|\{{\bf x}\in \mathbb{F}_{2}^{n} \ | \ f({\bf x})+g({\bf x})+h({\bf x})+{\bf w}\cdot {\bf x}=0 \}|\\
				&=2^{n}-2\text{wt}({\bf w})\\
				\text{wt}({\bf w})&=\frac{2^{n}-\widehat{f+g+h}({\bf w})}{2}
			\end{align*}
	All other cases follow in the same manner, which is similar to Theorem 3.1 in \cite{13}. Hence, we conclude the result.
		\end{proof}
	\end{theorem}
	
	\begin{theorem}\label{thm32}
		Let $\mathscr{C}_{f,g,h}$ be a binary code defined in (\ref{eq31}). Then $\mathscr{C}_{f,g,h}$ is minimal if and only if following two conditions hold simultaneously for any ${\bf x},{\bf y}\in \mathbb{F}_{2}^{n}$
		\begin{enumerate}
			\item For any $\phi_{1},\phi_{2}\in \mathcal{F}$ and ${\bf x}\neq {\bf y}$, we have
			\begin{equation*}
				\widehat{\phi_{1}}({\bf x})+\widehat{\phi_{2}}({\bf y})\neq 2^{n} \quad \text{and} \quad \widehat{\phi_{1}}({\bf x})-\widehat{\phi_{2}}({\bf y})\neq 2^{n}.
			\end{equation*}
			\item For any $\phi_{1},\phi_{2}\in \mathcal{F}$ with $\phi_{1}\neq\phi_{2}$, we have
			\begin{equation*}
				\widehat{\phi_{1}}({\bf x}+{\bf y})+\widehat{\phi_{2}}({\bf x})-\widehat{\phi_{1}+\phi_{2}}({\bf y})\neq 2^{n}
			\end{equation*}
		\end{enumerate}
		\begin{proof}
			First, we rewrite each codeword in $\mathscr{C}_{f,g,h}$ as follows,
			$c_{f,g,h}({\bf w})=\alpha{\bf f}+\beta{\bf g}+\gamma {\bf h}+c({\bf w})$, where ${\bf f}=(f({\bf x}))_{{\bf x}\in \mathbb{F}_{2}^{n*}}$, ${\bf g}=(g({\bf x}))_{{\bf x}\in \mathbb{F}_{2}^{n*}}$, ${\bf h}=(h({\bf x}))_{{\bf x}\in \mathbb{F}_{2}^{n*}}$ and $c({\bf w})\in \mathscr{C}=\{c({\bf w})=({\bf w}\cdot{\bf x})_{{\bf x}\in \mathbb{F}_{2}^{n*}}\ | \ {\bf w}\in \mathbb{F}_{2}^{n} \}$. Note that $\mathscr{C}$ is a simplex code with parameters $[2^{n}-1,n,2^{n-1}]$. Clearly, each codeword in $\mathscr{C}$ has weight $2^{n-1}$. Also, from the Theorem \ref{eq31}, we have, $\text{wt}({\bf f}+{\bf g}+c({\bf w}))=2^{n-1}-\frac{\widehat{f+g}({\bf w})}{2}$. We next consider the following cases for codewords in $\mathscr{C}_{f,g,h}$\\
			{\bf Case I:} If  $c_{1},c_{2}\in \{c({\bf w}),{\bf f}+c({\bf w}),{\bf g}+c({\bf w}),{\bf h}+c({\bf w}),{\bf f+g}+c({\bf w}),{\bf f+h}+c({\bf w}),{\bf g+h}+c({\bf w}),{\bf f+g+h}+c({\bf w})\}$, then consider the following subcases,\\
			(a) For $c_{1}=c({\bf w})$ and $c_{2}\in \{{\bf f}+c({\bf w}),{\bf g}+c({\bf w}),{\bf h}+c({\bf w}),{\bf f+g}+c({\bf w}),{\bf f+h}+c({\bf w}),{\bf g+h}+c({\bf w}),{\bf f+g+h}+c({\bf w})\}$. We give the proof in the case $c_{2}={\bf f+g+h}+c({\bf w})$, all other cases also follow from this. From Lemma \ref{lemma2.2} and Theorem \ref{eq31}, we get
			\begin{align*}
				c_{1}\preceq c_{2} \Longleftrightarrow&\text{wt}({\bf f+g+h})=\text{wt}({\bf f+g+h}+c({\bf w}))-2^{n-1}\\
				\Longleftrightarrow& \widehat{f+g+h}({\bf 0})-\widehat{f+g+h}({\bf w})=2^{n}
			\end{align*}
			and
			\begin{align*}
				c_{2}\preceq c_{1} \Longleftrightarrow&\text{wt}({\bf f+g+h})=2^{n-1}-\text{wt}({\bf f+g+h}+c({\bf w}))\\
				\Longleftrightarrow & \widehat{f+g+h}({\bf 0})+\widehat{f+g+h}({\bf w})=2^{n}
			\end{align*}
			(b) For $c_{1},c_{2}\in \{{\bf f}+c({\bf w}),{\bf g}+c({\bf w}),{\bf h}+c({\bf w}),{\bf f+g}+c({\bf w}),{\bf f+h}+c({\bf w}),{\bf g+h}+c({\bf w}),{\bf f+g+h}+c({\bf w})\} $ with $c_{1}\neq c_{2}$. In this case, we give the proof for the following subcases all other subcases follow in the same way\\
			(i) If $c_{1}={\bf f}+c({\bf w})$ and $c_{2}={\bf f+g}+c({\bf w})$, then
			\begin{align*}
				c_{1}\preceq c_{2} \Longleftrightarrow&\text{wt}({\bf g})=\text{wt}({\bf f+g}+c({\bf w}))-\text{wt}({\bf f}+c({\bf w}))\\
				\Longleftrightarrow& \widehat{g}({\bf 0})+\widehat{f}({\bf w})-\widehat{f+g}({\bf w})=2^{n}
			\end{align*}
			and
			\begin{align*}
				c_{2}\preceq c_{1} \Longleftrightarrow&\text{wt}({\bf g})=\text{wt}({\bf f}+c({\bf w}))-\text{wt}({\bf f+g}+c({\bf w}))\\
				\Longleftrightarrow & \widehat{g}({\bf 0})+\widehat{f+g}({\bf w})-\widehat{f}({\bf w})=2^{n}
			\end{align*}
			(ii) If $c_{1}={\bf f}+c({\bf w})$ and $c_{2}={\bf g+h}+c({\bf w})$, then
			\begin{align*}
				c_{1}\preceq c_{2} \Longleftrightarrow&\text{wt}({\bf f+g+h})=\text{wt}({\bf g+h}+c({\bf w}))-\text{wt}({\bf f}+c({\bf w}))\\
				\Longleftrightarrow& \widehat{f+g+h}({\bf 0})+\widehat{f}({\bf w})-\widehat{g+h}({\bf w})=2^{n}
			\end{align*}
			and
			\begin{align*}
				c_{2}\preceq c_{1} \Longleftrightarrow&\text{wt}({\bf f+g+h})=\text{wt}({\bf f}+c({\bf w}))-\text{wt}({\bf g+h}+c({\bf w}))\\
				\Longleftrightarrow & \widehat{f+g+h}({\bf 0})+\widehat{g+h}({\bf w})-\widehat{f}({\bf w})=2^{n}
			\end{align*}
			{\bf Case II:} If  $c_{1}\in \{c({\bf w_{1}}),{\bf f}+c({\bf w_{1}}),{\bf g}+c({\bf w_{1}}),{\bf h}+c({\bf w_{1}}),{\bf f+g}+c({\bf w_{1}}),{\bf f+h}+c({\bf w_{1}}),{\bf g+h}+c({\bf w_{1}}),{\bf f+g+h}+c({\bf w_{1}})\}$ and   $c_{2}\in \{c({\bf w_{2}}),{\bf f}+c({\bf w_{2}}),{\bf g}+c({\bf w_{2}}),{\bf h}+c({\bf w_{2}}),{\bf f+g}+c({\bf w_{2}}),{\bf f+h}+c({\bf w_{2}}),{\bf g+h}+c({\bf w_{2}}),{\bf f+g+h}+c({\bf w_{2}})\}$, where ${\bf w_{1}}\neq{\bf w_{2}}$ and both are non-zero, then consider the following subcases:\\
			(a) For $c_{1}=c({\bf w_{1}})$ and $c_{2}\in \{{\bf f}+c({\bf w_{2}}),{\bf g}+c({\bf w_{2}}),{\bf h}+c({\bf w_{2}}),{\bf f+g}+c({\bf w_{2}}),{\bf f+h}+c({\bf w_{2}}),{\bf g+h}+c({\bf w_{2}}),{\bf f+g+h}+c({\bf w_{2}})\}$, we give the proof in the case $c_{2}={\bf f+g+h}+c({\bf w_{2}})$, all other cases follow in the same manner. Now, we have
			\begin{align*}
				c_{1}\preceq c_{2} \Longleftrightarrow&\text{wt}({\bf f+g+h}+c({\bf w_{1}}+{\bf w_{2}}))=\text{wt}({\bf f+g+h}+c({\bf w_{2}}))-2^{n-1}\\
				\Longleftrightarrow& \widehat{f+g+h}({\bf w_{1}}+{\bf w_{2}})-\widehat{f+g+h}({\bf w_{2}})=2^{n}
			\end{align*}
			and
			\begin{align*}
				c_{2}\preceq c_{1} \Longleftrightarrow&\text{wt}({\bf f+g+h+c({\bf w_{1}}+{\bf w_{2}})})=2^{n-1}-\text{wt}({\bf f+g+h}+c({\bf w_{2}}))\\
				\Longleftrightarrow & \widehat{f+g+h}({\bf w_{1}}+{\bf w_{2}})+\widehat{f+g+h}({\bf w_{2}})=2^{n}
			\end{align*}
			(b) For $c_{1}={\bm\phi}+c({\bf w_{1}})$ and $c_{2}= {\bm\phi}+c({\bf w_{2}})$, where ${\bm \phi}\in \{{\bf f},{\bf g},{\bf h},{\bf f+g},{\bf f+h},{\bf g+h},{\bf f+g+h}\}$, we give the proof in the case ${\bm\phi}={\bf g+h}$, all other cases follow in the same manner. Now, we have
			\begin{align*}
				c_{1}\preceq c_{2} \Longleftrightarrow&\text{wt}(c({\bf w_{1}}+{\bf w_{2}}))=\text{wt}({\bf g+h}+c({\bf w_{2}}))-\text{wt}({\bf g+h}+c({\bf w_{1}}))\\
				\Longleftrightarrow& \widehat{g+h}({\bf w_{1}})-\widehat{g+h}({\bf w_{2}})=2^{n}
			\end{align*}
			and
			\begin{align*}
				c_{2}\preceq c_{1} \Longleftrightarrow&\text{wt}(c({\bf w_{1}}+{\bf w_{2}}))=\text{wt}({\bf g+h}+c({\bf w_{1}}))-\text{wt}({\bf g+h}+c({\bf w_{2}}))\\
				\Longleftrightarrow & \widehat{g+h}({\bf w_{2}})-\widehat{g+h}({\bf w_{1}})=2^{n}
			\end{align*}
			(c) For $c_{1}={\bm\phi_{1}}+c({\bf w_{1}})$ and $c_{2}= {\bm\phi_{2}}+c({\bf w_{2}})$,\\ where ${\bm \phi_{1}}, {\bm\phi_{2}}\in \{{\bf f},{\bf g},{\bf h},{\bf f+g},{\bf f+h},{\bf g+h},{\bf f+g+h}\}$, we give the proof in the case ${\bm\phi_{1}}={\bf f}$ and  ${\bm\phi_{2}}={\bf g+h}$, all other cases follow in the same manner. Now, we have
			\begin{align*}
				c_{1}\preceq c_{2} \Longleftrightarrow&\text{wt}({\bf f+g+h}+c({\bf w_{1}}+{\bf w_{2}}))=\text{wt}({\bf g+h}+c({\bf w_{2}}))-\text{wt}({\bf f}+c({\bf w_{1}}))\\
				\Longleftrightarrow& \widehat{f+g+h}({\bf w_{1}}+{\bf w_{2}})+\widehat{f}({\bf w_{1}})-\widehat{g+h}({\bf w_{2}})=2^{n}
			\end{align*}
			and
			\begin{align*}
				c_{2}\preceq c_{1} \Longleftrightarrow&\text{wt}({\bf f+g+h}+c({\bf w_{1}}+{\bf w_{2}}))=\text{wt}({\bf f}+c({\bf w_{1}}))-\text{wt}({\bf g+h}+c({\bf w_{2}}))\\
				\Longleftrightarrow & \widehat{f+g+h}({\bf w_{1}}+{\bf w_{2}})+\widehat{g+h}({\bf w_{2}})-\widehat{f}({\bf w_{1}})=2^{n}
			\end{align*}
		\end{proof}
	\end{theorem}
	\section{A family of Minimal Binary Linear Codes of dimension $n+3$}
In this section, with the help of code $\mathscr{C}_{f,g,h}$, we construct a family of minimal binary linear codes using a class of special Boolean functions. In particular, we provide a class of the  codes, which doesn't obey Ashikhmin-Barg condition.  \\
Throughout this section, we assume that $n$ is a positive even integer and $t=\frac{n}{2}$. A partial spread of order $\nu$ in $\mathbb{F}_{2}^{n}$ is a collection of $\nu$ pairwise disjoint t-dimensional subspaces $W_{1},W_{2},\ldots, W_{\nu}$. Also, observe that $\nu \leq 2^{t}+1$. Partial spread can be used to construct the Bent function\cite{3}. In a similar manner, we will use partial spread to obtain a class of special Boolean functions as defined below. \\
Let $f_{i}:\mathbb{F}_{2}^{n}\rightarrow \mathbb{F}_{2}$ be a Boolean function, and for $1\leq i\leq 2^{t}+1$, defined as
	\begin{equation*}
	f_{i}({\bf x})=\begin{cases}
1, \text{ if } {\bf x}\in W_{i}\setminus\{{\bf 0}\}\\
0,	\text{ Otherwise}
	\end{cases}
	\end{equation*}
For $1\leq s_{1},s_{2},s_{3}\leq 2^{t}+1$, let $\mathcal{A}_{1}=\{i_{1,}i_{2},\ldots, i_{s_{1}}\}, \mathcal{A}_{2}=\{j_{1},j_{2},\ldots, j_{s_{2}}\}$ and $\mathcal{A}_{3}=\{k_{1},k_{2},\ldots, k_{s_{3}}\}$. Define $f, g$  and $h$ as follows
\begin{equation}\label{eq41}
	f=\sum_{i\in \mathcal{A}_{1}}f_{i},\quad g=\sum_{i\in \mathcal{A}_{2}}f_{i} \quad \text{and} \quad \ h=\sum_{i\in \mathcal{A}_{3}}f_{i}
\end{equation}
We define $f+g, f+h,g+h$ and $f+g+h$ as follows
\begin{equation}\label{eq42}
	f+g=\sum_{i\in \mathcal{A}_{1}\Delta\mathcal{A}_{2}}f_{i},\quad f+h=\sum_{i\in \mathcal{A}_{1}\Delta\mathcal{A}_{3}}f_{i} \quad \quad g+h=\sum_{i\in \mathcal{A}_{2}\Delta\mathcal{A}_{3}}f_{i} \quad
	\text{and} \quad
	f+g+h=\sum_{i\in \mathcal{A}_{1}\Delta\mathcal{A}_{2}\Delta\mathcal{A}_{3}}f_{i}
\end{equation}
We use the following notation to denote the cardinality of the above defined set:\\
$|\mathcal{A}_{1}\Delta\mathcal{A}_{2}|=\chi_{12}=s_{1}+s_{2}-2s_{12}$\\ $|\mathcal{A}_{1}\Delta\mathcal{A}_{3}|=\chi_{13}=s_{1}+s_{3}-2s_{13},$\\ $|\mathcal{A}_{2}\Delta\mathcal{A}_{3}|=\chi_{23}=s_{2}+s_{3}-2s_{23}$\\
$|\mathcal{A}_{1}\Delta\mathcal{A}_{2}\Delta \mathcal{A}_{3}|=\chi_{123}=s_{1}+s_{2}+s_{3}-2s_{12}-2s_{13}-2s_{23}+4s_{123}$\\
With the above notations, we give the weight distribution of $\mathscr{C}_{f,g,h}$
	\begin{theorem}\label{thm41}
	Let the notations defined above, if $\mathcal{A}_{i}$ are distinct sets for $i\in \{1,2,3\}$ and $\mathcal{A}_{1}\Delta\mathcal{A}_{2}\Delta \mathcal{A}_{3}\neq \emptyset $, then $\mathscr{C}_{f,g,h}$ is a $[2^{n}-1,n+3]$ binary linear code whose weight distribution is given in Table \ref{table1}
		\begin{center}
			\begin{table}[h]
				\begin{tabular}{|c|c|c|c|}
					\hline
					Weight $w$ & Multiplicity $A_{W}$ & Weight $w$ & Multiplicity $A_{W}$\\
					\hline
					0& 1 & $2^{n-1}-s_{3}$ & $(2^{t}+1-s_{3})(2^{t}-1)$\\
					$s_{1}(2^{t}-1)$ & 1 & 	$2^{n-1}-\chi_{12}$ & $(2^{t}+1-\chi_{12})(2^{t}-1)$\\
					$s_{2}(2^{t}-1)$ & 1 & $2^{n-1}-\chi_{13}$ & $(2^{t}+1-\chi_{13})(2^{t}-1)$\\
					$s_{3}(2^{t}-1)$ & 1 & $2^{n-1}-\chi_{23}$ & $(2^{t}+1-\chi_{23})(2^{t}-1)$\\
					$\chi_{12}(2^{t}-1)$ & 1 & 	$2^{n-1}-\chi_{123}$ & $(2^{t}+1-\chi_{123})(2^{t}-1)$\\
					$\chi_{13}(2^{t}-1)$ & 1 & 	$2^{n-1}+2^{t}-s_{1}$ & $s_{1}(2^{t}-1)$\\
					$\chi_{23}(2^{t}-1)$ & 1 & $2^{n-1}+2^{t}-s_{2}$ & $s_{2}(2^{t}-1)$\\
					& & $2^{n-1}+2^{t}-s_{3}$ & $s_{3}(2^{t}-1)$\\
					$\chi_{123}(2^{t}-1)$ & 1 &  	$2^{n-1}+2^{t}-\chi_{12}$ & $\chi_{12}(2^{t}-1)$\\
					$2^{n-1}$ & $2^{n}-1$ & 	$2^{n-1}+2^{t}-\chi_{13}$ & $\chi_{13}(2^{t}-1)$\\
					$2^{n-1}-s_{1}$ & $(2^{t}+1-s_{1})(2^{t}-1)$ & $2^{n-1}+2^{t}-\chi_{23}$ & $\chi_{23}(2^{t}-1)$\\
					$2^{n-1}-s_{2}$ & $(2^{t}+1-s_{2})(2^{t}-1)$ & 	$2^{n-1}+2^{t}-\chi_{123}$ & $\chi_{123}(2^{t}-1)$\\	
					\hline
				\end{tabular}
				\caption{Weight Distribution of $\mathscr{C}_{f,g,h}$}
				\label{table1}
			\end{table}
		\end{center}
		\begin{proof}
			For any ${\bf w}\in \mathbb{F}_{2}^{n}$, we have three cases ${\bf w}={\bf 0}$ or ${\bf w}\neq {\bf 0}$ and ${\bf w}\notin W_{i}^{\perp} \ \forall \  i$ or ${\bf w}\in W^{\perp}_{i}$ for some $i$. From the definition of $f$ and $f_{i}$, we have
			\begin{align*}
				\tilde{f}({\bf 0})&=\sum_{{\bf x}\in \mathbb{F}^{n}_{2}}\sum_{i\in \mathcal{A}_{i}}f_{i}({\bf x})(-1)^{{\bf 0}\cdot {\bf x}}\\
				&=(|W_{i}|-1)+(|W_{i}|-1)+\cdots +(|W_{i}|-1)\\
				& = s_{1}(2^{t}-1)
			\end{align*}
			For ${\bf w}\notin W_{i}^{\perp} \ \forall i \in \mathcal{A}_{1}$. Note that, if ${\bf w}\notin W_{i}^{\perp} $, then ${\bf w}.{\bf x}=0$  for exactly half of the elements in $W_{i}$ and hence
			\begin{align*}
				\sum_{{\bf x}\in \mathbb{F}_{2}^{n}}f_{i}({\bf x})(-1)^{{\bf w}\cdot {\bf x}}&=-1\\
				\Rightarrow\tilde{f}({\bf w})&=\sum_{{\bf x}\in \mathbb{F}_{2}^{n}}\sum_{i\in \mathcal{A}_{1}}f_{i}({\bf x})(-1)^{{\bf w}\cdot {\bf x}}\\
				&= -s_{1}
			\end{align*}
			Similarly, we get $\tilde{f}({\bf w})=2^{t}-s_{1}$, if ${\bf w}\in W_{i}$, for some $i \in \mathcal{A}_{1}$. From Lemma \ref{lemma2.1}, we have
			\begin{equation*}
				\widehat{f}({\bf w})=
				\begin{cases}
					2^{n}-2s_{1}(2^{t}-1) & \text{if } {\bf w}=0\\
					2s_{1} & \text{if } {\bf w}\notin W_{i}^{\perp} \ \forall i\in \mathcal{A}_{1}\\
					-2^{t+1}+2s_{1} & \text{if } {\bf w}\neq{\bf 0} \text{ and } {\bf w}\in W_{i}^{\perp} \text{ for some } i\in \mathcal{A}_{1}
				\end{cases}
			\end{equation*}
		In a similar way, we find $\widehat{\phi}$, for all $\phi\in \mathcal{F}$ and the remaining proof follows from Theorem \ref{eq31}.
		\end{proof}
	\end{theorem}
According to Theorem \ref{thm31}, for the code $\mathscr{C}_{f,g,h}$ to be minimal, we impose the following conditions on $\mathcal{A}_{1},\mathcal{A}_{2}$ and $\mathcal{A}_{3}$.
\begin{enumerate}
	\item
		$\mathcal{A}_{k}\not\subset \mathcal{A}_{i}\Delta\mathcal{A}_{j} \quad \text{and} \quad \mathcal{A}_{i}\Delta\mathcal{A}_{j}\not\subset A_{k}, \quad \forall i,j,k\in \{1,2,3\}$
	
	\item For atleast two set in $\{\mathcal{A}_{1}\cap\mathcal{A}_{2},\mathcal{A}_{1}\cap\mathcal{A}_{3}, \mathcal{A}_{2}\cap\mathcal{A}_{3}\}$,\\ we have $\mathcal{A}_{1}\cap\mathcal{A}_{2}\cap \mathcal{A}_{3} \neq \mathcal{A}_{i}\cap\mathcal{A}_{j}$  and $\mathcal{A}_{1}\cap \mathcal{A}_{2}\cap\mathcal{A}_{3}\neq \phi$
	\item $\chi_{ij}=|\mathcal{A}_{i}\Delta\mathcal{A}_{j}|\geq 2$ and $\chi_{123}=|\mathcal{A}_{1}\Delta\mathcal{A}_{2}\Delta \mathcal{A}_{3}|\geq 2$
\end{enumerate}
Observe that these conditions are generalized version of the special case given in \cite{13}.\\
Following lemmas are immediate consequences of the above condition, which will be useful in proving $\mathscr{C}_{f,g,h}$ is a minimal linear code.
\begin{lemma}\label{prop1}
Let $\mathcal{A}_{i}$ are non-empty sets, for $i\in\{1,2,3\}$. If condition 1 is true for all $\mathcal{A}_{i}$, then following hold:
\begin{enumerate}
\item $\mathcal{A}_{i}\not\subset \mathcal{A}_{j}$ and $\mathcal{A}_{i}\cap \mathcal{A}_{j}\neq \phi$, for all $i,j$.
\item $(\mathcal{A}_{i}\cap\mathcal{A}_{j})\Delta(\mathcal{A}_{i}\cap\mathcal{A}_{k})\subsetneq \mathcal{A}_{j}\Delta\mathcal{A}_{k}$.
\end{enumerate}
\begin{proof}
\begin{enumerate}
	\item Suppose, $\mathcal{A}_{i}\subseteq\mathcal{A}_{j}$, then $\mathcal{A}_{i}\cup\mathcal{A}_{j}=\mathcal{A}_{j}$ and $\mathcal{A}_{i}\cap\mathcal{A}_{j}=\mathcal{A}_{i}$, this give, $\mathcal{A}_{i}\Delta\mathcal{A}_{j}\subseteq \mathcal{A}_{j}$, which is a contradiction.\\
	If $\mathcal{A}_{i}\cap\mathcal{A}_{j}=\phi$, then $\mathcal{A}_{i}\subseteq\mathcal{A}_{i}\Delta\mathcal{A}_{j}$, which is also not possible.
	\item If ${\bf x}\in (
	\mathcal{A}_{i}\cap \mathcal{A}_{j})\Delta(\mathcal{A}_{i}\cap\mathcal{A}_{k})$, which implies ${\bf x}\in \mathcal{A}_{i}\cap(\mathcal{A}_{j}\cup \mathcal{A}_{k})$ and ${\bf x}\notin \mathcal{A}_{i}\cap\mathcal{A}_{j}\cap\mathcal{A}_{k}$ , we have  $(\mathcal{A}_{i}\cap\mathcal{A}_{j})\Delta(\mathcal{A}_{i}\cap\mathcal{A}_{k})\subseteq \mathcal{A}_{j}\Delta\mathcal{A}_{k}$. Now, assume that, $(\mathcal{A}_{i}\cap\mathcal{A}_{j})\Delta(\mathcal{A}_{i}\cap\mathcal{A}_{k})=\mathcal{A}_{j}\Delta\mathcal{A}_{k}$.\\ Note that, $(\mathcal{A}_{i}\cap\mathcal{A}_{j})\Delta(\mathcal{A}_{i}\cap\mathcal{A}_{k})\subseteq \mathcal{A}_{i}$, which implies, $\mathcal{A}_{j}\Delta\mathcal{A}_{k}\subseteq A_{i}$. Which is a contradiction.
\end{enumerate}
\end{proof}
\end{lemma}
\begin{lemma}\label{prop2}
Let $\mathcal{A}_{i}$ are non-empty sets, for $i\in \{1,2,3\}$ and $2\leq s_{i}\leq 2^{t-1}$. If conditions 1 and 2 hold simultaneously, then $\chi_{ij}\leq 2^{t}-2$ and $\chi_{123}\leq 2^{t}-2$.
\begin{proof}
Since, $\chi_{ij}=s_{i}+s_{j}-2s_{ij}\leq 2^{t-1}+2^{t-1}-2=2^{t}-2$.\\
Note that, $|\mathcal{A}_{1}\cup \mathcal{A}_{2}\cup \mathcal{A}_{3}|\leq 2^{t}+1$. We know that,
\begin{align*}
\chi_{123}&=s_{1}+s_{2}+s_{3}-2s_{12}-2s_{13}-2s_{23}+4s_{123}\\
&=s_{1}+s_{2}+s_{3}-s_{12}-s_{13}-s_{23}+2s_{123}-(s_{12}+s_{13}+s_{23}-2s_{123})\\
&\leq 2^{t}+1-(s_{12}+s_{13}+s_{23}-2s_{123})\\
&\leq 2^{t}+1-3=2^{t}-2
\end{align*}
\end{proof}
\end{lemma}
 From onwards, we assume that sets $\mathcal{A}_{1},\mathcal{A}_{2}$ and $\mathcal{A}_{3}$ satisfy the above stated conditions, we also give the  values of Walsh transforms for $\phi \in \mathcal{F}$ in Table \ref{table2}.
	\begin{table}[h]
		\begin{center}
		\begin{tabular}{|p{3.5cm}|c|c|c|}
			\hline
			Walsh
 Transform/ Condition & ${\bf w}={\bf 0}$ & ${\bf w}\notin W^{\perp}_{i}, \ \forall i\in \mathcal{A}$ & ${\bf w}\in W^{\perp}_{i}$ for some $i \in \mathcal{A}$  \\
			\hline
			$\widehat{f}({\bf w})$ & $2^{n}-2s_{1}(2^{t}-1)$ & $2s_{1}$ & $-2^{t+1}+2s_{1}$ \\
			\hline
			$\widehat{g}({\bf w})$ & $2^{n}-2s_{2}(2^{t}-1)$ & $2s_{2}$ & $-2^{t+1}+2s_{2}$ \\
			\hline
			$\widehat{h}({\bf w})$ & $2^{n}-2s_{3}(2^{t}-1)$ & $2s_{3}$ & $-2^{t+1}+2s_{3}$ \\
			\hline
			$\widehat{f+g}({\bf w})$ & $2^{n}-2\chi_{12}(2^{t}-1)$ & $2\chi_{12}$ & $-2^{t+1}+2\chi_{12}$ \\
			\hline
			$\widehat{f+h}({\bf w})$ & $2^{n}-2\chi_{13}(2^{t}-1)$ & $2\chi_{13}$ & $-2^{t+1}+2\chi_{13}$ \\
			\hline
			 $\widehat{g+h}({\bf w})$ & $2^{n}-2\chi_{23}(2^{t}-1)$ & $2\chi_{23}$ & $-2^{t+1}+2\chi_{23}$ \\
			\hline
			$\widehat{f+g+h}({\bf w})$ & $2^{n}-2\chi_{123}(2^{t}-1)$ & $2\chi_{123}$ & $-2^{t+1}+2\chi_{123}$ \\
			\hline
		\end{tabular}
	\end{center}
		\caption{The values of Walsh Transforms}
		\label{table2}
	\end{table}

	\begin{proposition}\label{lemma41}
	Let $n\geq 6$ and $2\leq s\leq 2^{t-1}$, for $s\in \{s_{1},s_{2},s_{3}\}$ and $\mathcal{F}$ be a collection of Boolean functions defined in \ref{eq41} and \ref{eq42}. Then, for ${\bf x},{\bf y}\in \mathbb{F}_{2}^{n}$ with ${\bf x}\neq {\bf y}$, we have
		\begin{equation*}
			\widehat{\phi}_{1}({\bf x})+\widehat{\phi}_{2}({\bf y})\neq 2^{n}
		\end{equation*}
		for all, $\phi_{1},\phi_{2}\in \mathcal{F}$.
		\begin{proof} We can observe various cases by considering different values of $\phi_{1}$ and $\phi_{2}$. Within each case, there are seven subcases that arise from different choices of ${\bf x}$ and ${\bf y}$. We discuss the proof for two subcases of each possibility.\\
			{\bf Case I:} For $\phi_{1}=\phi_{2}=f$, we have\\
			(a) If ${\bf x}={\bf 0}$ and ${\bf y}\notin W_{i}^{\perp}$, for all $i\in \mathcal{A}_{1}$.
			\begin{align*}
				&\widehat{f}({\bf 0})+\widehat{f}({\bf y})\neq 2^{n} \Leftrightarrow 2^{n}-2s_{1}(2^{t}-1)+2s_{1}\neq 2^{n}\Leftrightarrow s_{1}(2^{t}-2)\neq 0
			\end{align*}
			(b) If ${\bf x}=0$ and ${\bf y}\in W_{i}^{\perp}$, for some $i\in \mathcal{A}_{1}$.
			\begin{align*}
				&\widehat{f}({\bf 0})+\widehat{f}({\bf y})\neq 2^{n} \Leftrightarrow 2^{n}-2s_{1}(2^{t}-1)-2^{t+1}+2s_{1}\neq 2^{n}\Leftrightarrow s_{1}(2^{t}-2)\neq -2^{t}
			\end{align*}
	Proof of $\phi_{1}=\phi_{2}=f+g$ and $\phi_{1}=\phi_{2}=f+g+h$, follows from {\bf Case I} and Lemma \ref{prop2}.\\
	{\bf Case II:} For $\phi_{1}=f$ and $\phi_{2}=g$, we have\\
	(a) If ${\bf x}={\bf 0}$ and ${\bf y}\notin W_{i}^{\perp}$, for all $i\in \mathcal{A}_{2}$.
			\begin{align*}
				&\widehat{f}({\bf 0})+\widehat{g}({\bf y})\neq 2^{n} \Leftrightarrow 2^{n}-2s_{1}(2^{t}-1)+2s_{2}\neq 2^{n}\Leftrightarrow s_{2}\neq s_{1}(2^{t}-1)
			\end{align*}
			(b) If ${\bf x}=0$ and ${\bf y}\in W_{i}^{\perp}$, for some $i\in \mathcal{A}_{2}$.
			\begin{align*}
				&\widehat{f}({\bf 0})+\widehat{g}({\bf y})\neq 2^{n} \Leftrightarrow 2^{n}-2s_{1}(2^{t}-1)-2^{t+1}+2s_{2}\neq 2^{n}
				\Leftrightarrow s_{1}+s_{2}\neq 2^{t}s_{1}+2^{t}
			\end{align*}
	{\bf Case III:} For $\phi_{1}=f$ and $\phi_{2}=f+g$, we have\\
	(a) If ${\bf x}={\bf 0}$ and ${\bf y}\notin W_{i}^{\perp}$, for all $i\in \mathcal{A}_{1}\Delta\mathcal{A}_{2}$.
	\begin{align*}
		&\widehat{f}({\bf 0})+\widehat{f+g}({\bf y})\neq 2^{n} \Leftrightarrow 2^{n}-2s_{1}(2^{t}-1)+2\chi_{12}\neq 2^{n}\Leftrightarrow \chi_{12}\neq s_{1}(2^{t}-1)
	\end{align*}
	(b) If ${\bf y}={\bf 0}$ and ${\bf x}\notin W_{i}^{\perp}$, for all $i\in \mathcal{A}_{1}$.
	\begin{align*}
		&\widehat{f}({\bf x})+\widehat{f+g}({\bf 0})\neq 2^{n} \Leftrightarrow 2s_{1}+2^{n}-2\chi_{12}(2^{t}-1)\neq 2^{n}
		\Leftrightarrow s_{1}\neq \chi_{12}(2^{t}-1)
	\end{align*}
{\bf Case IV:} If $\phi_{1}=f$ and $\phi_{2}=g+h$, then this case follows from {\bf Case III}.\\
{\bf Case V:} For $\phi_{1}=f$ and $\phi_{2}=f+g+h$, we have\\
(a) If ${\bf x}={\bf 0}$ and ${\bf y}\notin W_{i}^{\perp}$, for all $i\in \mathcal{A}_{1}\Delta\mathcal{A}_{2}\Delta\mathcal{A}_{3}$.
\begin{align*}
	&\widehat{f}({\bf 0})+\widehat{f+g+h}({\bf y})\neq 2^{n} \Leftrightarrow 2\chi_{123}+2^{n}-2s_{1}(2^{t}-1)\neq 2^{n}\Leftrightarrow
	\chi_{123}\neq s_{1}(2^{t}-1)
\end{align*}
(b) If ${\bf y}={\bf 0}$ and ${\bf x}\notin W_{i}^{\perp}$, for all $i\in \mathcal{A}_{1}$.
\begin{align*}
	&\widehat{f}({\bf x})+\widehat{f+g+h}({\bf 0})\neq 2^{n} \Leftrightarrow 2s_{1}+2^{n}-2\chi_{123}(2^{t}-1)\neq 2^{n}
 \Leftrightarrow s_{1}\neq \chi_{123}(2^{t}-1)
\end{align*}
{\bf Case VI:} For $\phi_{1}=f+g$ and $\phi_{2}=f+g+h$, we have\\
(a) If ${\bf x}={\bf 0}$ and ${\bf y}\notin W_{i}^{\perp}$, for all $i\in \mathcal{A}_{1}\Delta\mathcal{A}_{2}\Delta\mathcal{A}_{3}$.
\begin{align*}
	&\widehat{f+g}({\bf 0})+\widehat{f+g+h}({\bf y})\neq 2^{n}\Leftrightarrow 2\chi_{123}+2^{n}-2\chi_{12}(2^{t}-1)\neq 2^{n}
	\Leftrightarrow \chi_{123}\neq \chi_{12}(2^{t}-1)
\end{align*}
(b) If ${\bf y}={\bf 0}$ and ${\bf x}\in W_{i}^{\perp}$, for some $i\in \mathcal{A}_{1}\Delta\mathcal{A}_{2}$.
\begin{align*}
	&\widehat{f+g}({\bf x})+\widehat{f+g+h}({\bf 0})\neq 2^{n}\Leftrightarrow
	2\chi_{12}+2^{n}-2\chi_{123}(2^{t}-1)\neq 2^{n}\Leftrightarrow \chi_{12}\neq \chi_{123}(2^{t}-1)
\end{align*}
All other cases follow in the similar manner. Hence, $	\widehat{\phi}_{1}({\bf x})+\widehat{\phi}_{2}({\bf y})\neq 2^{n}$, for all $\phi_{1},\phi_{2}\in \mathcal{F}$ and ${\bf x},{\bf y}\in \mathbb{F}_{2}^{n}$, with ${\bf x}\neq {\bf y}$.
\end{proof}
\end{proposition}

	\begin{proposition}\label{lemma46}
		Let $n\geq 6$ and $2\leq s\leq 2^{t-1}$, for $s\in \{s_{1},s_{2},s_{3}\}$ and $\mathcal{F}$ be a collection of Boolean functions defined in \ref{eq41} and \ref{eq42}. Then, for ${\bf x},{\bf y}\in \mathbb{F}_{2}^{n}$ with ${\bf x}\neq {\bf y}$, we have
	\begin{equation*}
\widehat{\phi}_{1}({\bf x})-\widehat{\phi}_{2}({\bf y})\neq 2^{n}
	\end{equation*}
	for all, $\phi_{1},\phi_{2}\in \mathcal{F}$.
	\begin{proof} We can observe various cases by considering different values of $\phi_{1}$ and $\phi_{2}$. Within each case, there are seven subcases that arise from different choices of ${\bf x}$ and ${\bf y}$. We discuss the proof for few subcases for each possinilities.\\
		{\bf Case I:} For $\phi_{1}=\phi_{2}=f$, consider following subcases:\\
		(a) If ${\bf x}={\bf 0}$ and ${\bf y}\notin W_{i}^{\perp}$, for all $i\in \mathcal{A}_{1}$, then 
		\begin{align*}
			&   \widehat{f}({\bf 0})-\widehat{f}({\bf y})\neq 2^{n} \Leftrightarrow 2^{n}-2s_{1}(2^{t}-1)-2s_{1}\neq 2^{n}\Leftrightarrow 2^{t+1}s_{1}\neq 0\\
			\text{and}\\
			&\widehat{f}({\bf y})-\widehat{f}({\bf 0})\neq 2^{n} \Leftrightarrow 2s_{1}-2^{n}+2s_{1}(2^{t}-1)\neq 2^{n}\Leftrightarrow 2^{t+1}s_{1}\neq 2^{n+1}\Leftrightarrow s_{1}\neq2^{t}
		\end{align*}
		(b) If ${\bf x}={\bf 0}$ and ${\bf y}\in W_{i}^{\perp}$, for some $i\in \mathcal{A}_{1}$, then
		\begin{align*}
		&\widehat{f}({\bf 0})-\widehat{f}({\bf y})\neq 2^{n} \Leftrightarrow 2^{n}-2s_{1}(2^{t}-1)+2^{t+1}-2s_{1}\neq 2^{n}\Leftrightarrow -2s_{1}2^{t}+2^{t+1}\neq 0\Leftrightarrow s_{1}\neq 1\\
			\text{and}\\
			&\widehat{f}({\bf y})-\widehat{f}({\bf 0})\neq 2^{n} \Leftrightarrow -2^{t+1}+2s_{1}-2^{n}+2s_{1}(2^{t}-1)\neq 2^{n}\Leftrightarrow s_{1}\neq2^{t}+1
		\end{align*}
		Proof of $\phi_{1}=\phi_{2}=f+g$ and $\phi_{1}=\phi_{2}=f+g+h$, follows from {\bf Case I} and Lemma \ref{prop2}.\\
		{\bf Case II:} For $\phi_{1}=f$ and $\phi_{2}=g$, consider following subcases:\\
		(a) If ${\bf x}={\bf 0}$ and ${\bf y}\notin W_{i}^{\perp}$, for all $i\in \mathcal{A}_{2}$, then
		\begin{align*}
		&\widehat{f}({\bf 0})-\widehat{g}({\bf y})\neq 2^{n} \Leftrightarrow 2^{n}-2s_{1}(2^{t}-1)-2s_{2}\neq 2^{n}\Leftrightarrow s_{2}\neq -s_{1}(2^{t}-1)
		\end{align*}
		(b) If ${\bf x}={\bf 0}$ and ${\bf y}\in W_{i}^{\perp}$, for some $i\in \mathcal{A}_{2}$, then
		\begin{align*}
			&\widehat{f}({\bf 0})-\widehat{g}({\bf y})\neq 2^{n} \Leftrightarrow 2^{n}-2s_{1}(2^{t}-1)+2^{t+1}-2s_{2}\neq 2^{n}\Leftrightarrow s_{1}-s_{2}\neq 2^{t}(s_{1}-1)
		\end{align*}
		(c) If ${\bf x}\notin W_{i}$, for all $i\in \mathcal{A}_{1}$ and ${\bf y}={\bf 0}$, then
		\begin{align*}
			&\widehat{f}({\bf x})-\widehat{g}({\bf 0})\neq 2^{n} \Leftrightarrow 2s_{1}-2^{n}+2s_{2}(2^{t}-1)\neq 2^{n}\Leftrightarrow s_{1}+s_{2}(2^{t}-1)\neq 2^{n}
		\end{align*}
		(d) If ${\bf x}\in W_{i}$, for some $i\in \mathcal{A}_{1}$ and ${\bf y}={\bf 0}$, then
		\begin{align*}
			&\widehat{f}({\bf x})-\widehat{g}({\bf 0})\neq 2^{n} \Leftrightarrow -2^{t+1}+2s_{1}-2^{n}+2s_{2}(2^{t}-1)\neq 2^{n}\Leftrightarrow s_{1}+s_{2}(2^{t}-1)\neq 2^{n}+2^{t}
		\end{align*}
	{\bf Case III:} If $\phi_{1}=g$ and $\phi_{2}=f$, then this case follows from {\bf Case II}.\\
	{\bf Case IV:} For $\phi_{1}=f$ and $\phi_{2}=f+g$, consider following subcases:\\
		(a) If ${\bf x}={\bf 0}$ and ${\bf y}\notin W_{i}^{\perp}$, for all $i\in \mathcal{A}_{1}\Delta\mathcal{A}_{2}$, then
		\begin{align*}
			&\widehat{f}({\bf 0})-\widehat{f+g}({\bf y})\neq 2^{n} \Leftrightarrow 2^{n}-2s_{1}(2^{t}-1)-2\chi_{12}\neq 2^{n}\Leftrightarrow \chi_{12}\neq -s_{1}(2^{t}-1)
		\end{align*}
		(b) If ${\bf x}={\bf 0}$ and ${\bf y}\in W_{i}^{\perp}$, for some $i\in \mathcal{A}_{1}\Delta\mathcal{A}_{2}$, then
		\begin{align*}
			&\widehat{f}({\bf 0})-\widehat{f+g}({\bf y})\neq 2^{n} \Leftrightarrow 2^{n}-2s_{1}(2^{t}-1)+2^{t+1}-2\chi_{12}\neq 2^{n}\Leftrightarrow s_{1}(2^{t}-1)+\chi_{12}\neq 2^{t}
		\end{align*}
	(c) If ${\bf y}={\bf 0}$ and ${\bf x}\notin W_{i}^{\perp}$, for all $i\in \mathcal{A}_{1}$, then
	\begin{align*}
		&\widehat{f}({\bf x})-\widehat{f+g}({\bf 0})\neq 2^{n} \Leftrightarrow 2s_{1}-2^{n}+2\chi_{12}(2^{t}-1)\neq 2^{n}\Leftrightarrow s_{1}+\chi_{12}(2^{t}-1)\neq 2^{n}
	\end{align*}
	(d) If ${\bf y}={\bf 0}$ and ${\bf x}\in W_{i}^{\perp}$, for some $i\in \mathcal{A}_{1}$, then
	\begin{align*}
		&\widehat{f}({\bf x})-\widehat{f+g}({\bf 0})\neq 2^{n} \Leftrightarrow -2^{t+1}+2s_{1}-2^{n}+2\chi_{12}(2^{t}-1)\neq 2^{n}\Leftrightarrow s_{1}+\chi_{12}(2^{t}-1)\neq 2^{n}+2^{t}
	\end{align*}
{\bf Case V:} If $\phi_{1}=f+g$ and $\phi_{2}=f$, then this case follow from above.\\
{\bf Case VI:} For $\phi_{1}=f$ and $\phi_{2}=g+h$, consider following subcases:\\
		(a) If ${\bf x}={\bf 0}$ and ${\bf y}\notin W_{i}^{\perp}$, for all $i\in \mathcal{A}_{2}\Delta\mathcal{A}_{3}$, then
		\begin{align*}
			&\widehat{f}({\bf 0})-\widehat{g+h}({\bf y})\neq 2^{n} \Leftrightarrow 2^{n}-2s_{1}(2^{t}-1)-2\chi_{23}\neq 2^{n}\Leftrightarrow \chi_{23}\neq -s_{1}(2^{t}-1)
		\end{align*}
		(b) If ${\bf x}={\bf 0}$ and ${\bf y}\in W_{i}^{\perp}$, for some $i\in \mathcal{A}_{2}\Delta\mathcal{A}_{3}$, then
		\begin{align*}
			&\widehat{f}({\bf 0})-\widehat{g+h}({\bf y})\neq 2^{n} \Leftrightarrow 2^{n}-2s_{1}(2^{t}-1)+2^{t+1}-2\chi_{23}\neq 2^{n}\Leftrightarrow s_{1}(2^{t}-1)+\chi_{23}\neq 2^{t}
		\end{align*}
	{\bf Case VII:} If $\phi_{1}=g+h$ and $\phi_{2}=f$, then this case follow from above.\\
 	{\bf Case VIII:} For $\phi_{1}=f$ and $\phi_{2}=f+g+h$, consider following subcases:\\
		(a) If ${\bf x}={\bf 0}$ and ${\bf y}\notin W_{i}^{\perp}$, for all $i\in \mathcal{A}_{1}\Delta\mathcal{A}_{2}\Delta\mathcal{A}_{3}$, then
		\begin{align*}
			&\widehat{f}({\bf 0})-\widehat{f+g+h}({\bf y})\neq 2^{n} \Leftrightarrow 2^{n}-s_{1}(2^{t}-1)-2\chi_{123}\neq 2^{n}\Leftrightarrow \chi_{123}\neq -s_{1}(2^{t}-1)
		\end{align*}
		(b) If ${\bf x}={\bf 0}$ and ${\bf y}\in W_{i}^{\perp}$, for some $i\in \mathcal{A}_{1}\Delta\mathcal{A}_{2}\Delta\mathcal{A}_{3}$, then
		\begin{align*}
			&\widehat{f}({\bf 0})-\widehat{f+g+h}({\bf y})\neq 2^{n} \Leftrightarrow 2^{n}-s_{1}(2^{t}-1)+2^{t+1}-2\chi_{123}\neq 2^{n}\Leftrightarrow s_{1}(2^{t}-1)+\chi_{123}\neq 2^{t}
		\end{align*}
		{\bf Case IX:} For $\phi_{1}=f+g$ and $\phi_{2}=f+g+h$, consider following subcases:\\
		(a) If ${\bf x}={\bf 0}$ and ${\bf y}\notin W_{i}^{\perp}$, for all $i\in \mathcal{A}_{1}\Delta\mathcal{A}_{2}\Delta\mathcal{A}_{3}$, then
		\begin{align*}
			&\widehat{f+g}({\bf 0})-\widehat{f+g+h}({\bf y})\neq 2^{n}\Leftrightarrow 2^{n}-\chi_{12}(2^{t}-1)-2\chi_{123}\neq 2^{n}\Leftrightarrow \chi_{123}\neq -\chi_{12}(2^{t}-1)
		\end{align*}
		(b) If ${\bf x}={\bf 0}$ and ${\bf y}\in W_{i}^{\perp}$, for some $i\in \mathcal{A}_{1}\Delta\mathcal{A}_{2}\Delta\mathcal{A}_{3}$, then
		\begin{align*}
			&\widehat{f+g}({\bf 0})-\widehat{f+g+h}({\bf y})\neq 2^{n}\Leftrightarrow 2^{n}-\chi_{12}(2^{t}-1)+2^{t+1}-2\chi_{123}\neq 2^{n}\Leftrightarrow \chi_{12}(2^{t}-1)+\chi_{123}\neq 2^{t}
		\end{align*}
		All other cases follow in the similar manner. Hence, $\widehat{\phi_{1}}({\bf x})-\widehat{\phi_{2}}({\bf y})\neq 2^{n}$, for all $\phi_{1},\phi_{2}\in \mathcal{F}$ and ${\bf x},{\bf y}\in \mathbb{F}_{2}^{n}$ with ${\bf x}\neq {\bf y}$.
	\end{proof}
\end{proposition}

\begin{proposition}\label{lemma42}
	Let $n\geq 6$ and $2\leq s\leq 2^{t-1}$, for $s\in \{s_{1},s_{2},s_{3}\}$ and $\mathcal{F}$ be a collection of Boolean functions defined in \ref{eq41} and \ref{eq42}. Then, for ${\bf x},{\bf y}\in \mathbb{F}_{2}^{n}$ with ${\bf x}\neq {\bf y}$, we have
	\begin{equation*}
		\widehat{\phi}_{1}({\bf 0})+\widehat{\phi}_{2}({\bf x})-\widehat{(\phi_{1}+\phi_{2})}({\bf x})\neq 2^{n}
	\end{equation*}
	for all, $\phi_{1},\phi_{2}\in \mathcal{F}$ with $\phi_{1}\neq \phi_{2}$.
\begin{proof}
 For different choices of functions, we have following cases:\\
{\bf Case I:} If $\phi_{1}=f$ and $\phi_{2}=g$, then for ${\bf x}$, we have following subcases:\\
(a) If ${\bf x}={\bf 0}$, then
\begin{align*}
&\widehat{f}({\bf 0})+\widehat{g}({\bf 0})-\widehat{f+g}({\bf 0})\neq 2^{n}\\
&\Leftrightarrow 2^{n}-2s_{1}(2^{t}-1)+2^{n}-2s_{2}(2^{t}-1)-2^{n}+2\chi_{12}(2^{t}-1)\neq 2^{n} \Leftrightarrow s_{12}\neq 0
\end{align*}
(b) If ${\bf x}\notin W_{i}^{\perp}$, for all $i\in \mathcal{A}_{2}\cup(\mathcal{A}_{1}\Delta\mathcal{A}_{2})$, then
\begin{align*}
&\widehat{f}({\bf 0})+\widehat{g}({\bf x})-\widehat{f+g}({\bf x})\neq 2^{n}
\Leftrightarrow 2^{n}-2s_{1}(2^{t}-1)+2s_{2}-2\chi_{12}\neq 2^{n} \Leftrightarrow s_{12}\neq 2^{t-1}s_{1}
\end{align*}
(c) If ${\bf x}\in W_{i}^{\perp}$ for some $i\in \mathcal{A}_{2}\cup(\mathcal{A}_{1}\Delta\mathcal{A}_{2})$, then there are three subcases, we give proof for only one subcase and the other two cases follow from this.\\
If $i\notin \mathcal{A}_{2}$ and $i\in \mathcal{A}_{1}\Delta\mathcal{A}_{2}$, then
\begin{align*}
	&\widehat{f}({\bf 0})+\widehat{g}({\bf x})-\widehat{f+g}({
	\bf x})\neq 2^{n}\Leftrightarrow 2^{n}-2s_{1}(2^{t}-1)+2s_{2}+2^{t+1}-2\chi_{12}\neq 2^{n}\\
	& -s_{1}2^{t}+2^{t}+2s_{12}\neq 0\Leftrightarrow s_{12}\neq 2^{t-1}(s_{1}-1)
\end{align*}
{\bf Case II:} If $\phi_{1}=f$ and $\phi_{2}=f+g$, then for ${\bf x}$, we have following subcases:\\
(a) If ${\bf x}={\bf 0}$, then
\begin{align*}
	&\widehat{f}({\bf 0})+\widehat{f+g}({\bf 0})-\widehat{g}({\bf 0})\neq 2^{n}\\
	&\Leftrightarrow 2^{n}-2s_{1}(2^{t}-1)+2^{n}-2\chi_{12}(2^{t}-1)-2^{n}+2s_{2}(2^{t}-1)\neq 2^{n} \Leftrightarrow s_{12}\neq s_{1}
\end{align*}
(b) If ${\bf x}\notin W_{i}^{\perp}$, for all $i\in \mathcal{A}_{2}\cup(\mathcal{A}_{1}\Delta\mathcal{A}_{2})$, then
\begin{align*}
	&\widehat{f}({\bf 0})+\widehat{f+g}({\bf x})-\widehat{g}({\bf x})\neq 2^{n}
	\Leftrightarrow 2^{n}-2s_{1}(2^{t}-1)+2\chi_{12}-2s_{2}\neq 2^{n} \Leftrightarrow s_{12}\neq -s_{1}(2^{t-1}-1)
\end{align*}
(c) If ${\bf x}\in W_{i}^{\perp}$ for some $i\in\mathcal{A}_{2}\cup(\mathcal{A}_{1}\Delta\mathcal{A}_{2})$, then there are three subcases, we give proof for only one subcase and the other two cases follow from this.\\
If $i\in\mathcal{A}_{2}$ and $i\notin\mathcal{A}_{1}\Delta\mathcal{A}_{2}$, then
\begin{align*}
	&\widehat{f}({\bf 0})+\widehat{f+g}({\bf x})-\widehat{g}({\bf x})\neq 2^{n}\Leftrightarrow 2^{n}-2s_{1}(2^{t}-1)+2\chi_{12}+2^{t+1}-2s_{2}\neq 2^{n}\\
	& -s_{1}2^{t-1}+s_{1}-s_{12}+2^{t-1}\neq 0\Leftrightarrow s_{1}-s_{12}\neq 2^{t-1}(s_{1}-1)
\end{align*}
{\bf Case III:} If $\phi_{1}=f+g$ and $\phi_{2}=f$, then for ${\bf x}$, we have following subcases:\\
(a) If ${\bf x}={\bf 0}$, then this follows from above.\\
(b) If ${\bf x}\notin W_{i}^{\perp}$, for all $i\in \mathcal{A}_{1}\cup\mathcal{A}_{2}$, then
\begin{align*}
	&\widehat{f+g}({\bf 0})+\widehat{f}({\bf x})-\widehat{g}({\bf x})\neq 2^{n}
	\Leftrightarrow 2^{n}-2\chi_{12}(2^{t}-1)+2s_{1}-2s_{2}\neq 2^{n} \Leftrightarrow s_{1}-s_{12}\neq 2^{t-1}\chi_{12}
\end{align*}
(c)If ${\bf x}\in W_{i}^{\perp}$ for some $i\in\mathcal{A}_{1}\cup \mathcal{A}_{2}$,  then there are three subcases, we give proof for only one subcase and the other two cases follow from this.\\
If $i\in \mathcal{A}_{2}$ and $i\notin \mathcal{A}_{1}$, then
\begin{align*}
	&\widehat{f+g}({\bf 0})+\widehat{f}({\bf x})-\widehat{g}(x)\neq 2^{n}\Leftrightarrow 2^{n}-2\chi_{12}(2^{t}-1)+2s_{1}+2^{t+1}-2s_{2}\neq 2^{n}\\
	&\Leftrightarrow  s_{1}-s_{12}\neq 2^{t-1}(\chi_{12}-1)
\end{align*}
{\bf Case IV:} If $\phi_{1}=f$ and $\phi_{2}=g+h$, then for ${\bf x}$, we have following possibilities.\\
(a) If ${\bf x}={\bf 0}$, then
\begin{align*}
	&\widehat{f}({\bf 0})+\widehat{g+h}({\bf 0})-\widehat{f+g+h}({\bf 0})\neq 2^{n}\Leftrightarrow 2^{n}-2s_{1}(2^{t}-1)+2^{n}-2\chi_{23}(2^{t}-1)\\
	&-2^{n}+2\chi_{123}(2^{t}-1)\neq 2^{n} \Leftrightarrow(2^{t}-1)(-2s_{12}-2s_{13}+4s_{123})\neq 0 \Leftrightarrow 2s_{123}\neq s_{12}+s_{13}
\end{align*}
(b) If ${\bf x}\notin W_{i}^{\perp}$, for all $i\in (\mathcal{A}_{2}\Delta\mathcal{A}_{3})\cup(\mathcal{A}_{1}\Delta\mathcal{A}_{2}\Delta\mathcal{A}_{3})$, then
\begin{align*}
	&\widehat{f}({\bf 0})+\widehat{g+h}({\bf x})-\widehat{f+g+h}({\bf x})\neq 2^{n}\Leftrightarrow 2^{n}-2s_{1}(2^{t}-1)+2\chi_{23}-2\chi_{123}\neq 2^{n}\\
	& \Leftrightarrow -s_{1}(2^{t}-1)-(s_{1}-2s_{13}-2s_{13}+4s_{123})\neq 0 \Leftrightarrow s_{12}+s_{13}-2s_{123}\neq 2^{t-1}s_{1}
\end{align*}
(c) If ${\bf x}\in W_{i}^{\perp}$ for some $i\in(\mathcal{A}_{2}\Delta\mathcal{A}_{3})\cup(\mathcal{A}_{1}\Delta\mathcal{A}_{2}\Delta\mathcal{A}_{3})$, then there are three subcases, we give proof for only one subcase and the other two cases follow from this.\\
If $i\notin \mathcal{A}_{2}\Delta\mathcal{A}_{3}$ and $i\in \mathcal{A}_{1}\Delta\mathcal{A}_{2}\Delta\mathcal{A}_{3}$, then
\begin{align*}
	&\widehat{f}({\bf 0})+\widehat{g+h}({\bf x})-\widehat{f+g+h}({\bf x})\neq 2^{n}\Leftrightarrow 2^{n}-2s_{1}(2^{t}-1)+2\chi_{23}+2^{t+1}-2\chi_{123}\neq 2^{n}\\
	& \Leftrightarrow -s_{1}(2^{t}-1)+2^{t}-(s_{1}-2s_{13}-2s_{13}+4s_{123})\neq 0\Leftrightarrow  s_{12}+s_{13}-2s_{123}\neq 2^{t-1}(s_{1}-1)
\end{align*}
{\bf Case V:} If $\phi_{1}=g+h$ and $\phi_{2}=f$, then for ${\bf x}$, we have following possibilities.\\
(a) If ${\bf x}={\bf 0}$, then this follows from above.\\
(b) If ${\bf x}\notin E_{i}^{\perp}$, for all $i\in \mathcal{A}_{1}\cup(\mathcal{A}_{1}\Delta\mathcal{A}_{2}\Delta
\mathcal{A}_{3})$, then
\begin{align*}
	&\widehat{g+h}({\bf 0})+\widehat{f}({\bf x})-\widehat{f+g+h}({\bf x})\neq 2^{n}\Leftrightarrow 2^{n}-2\chi_{23}(2^{t}-1)+2s_{1}-2\chi_{123}\neq 2^{n}\\
	&\Leftrightarrow -2^{t}\chi_{23}-(-2s_{12}-2s_{13}+4s_{123})\neq 0
	\Leftrightarrow 2^{t-1}\chi_{23}\neq(s_{12}+s_{13}-2s_{123})
\end{align*}
(c) If ${\bf x}\in W_{i}^{\perp}$ for some $i\in\mathcal{A}_{1}\cup(\mathcal{A}_{1}\Delta\mathcal{A}_{2}\Delta
\mathcal{A}_{3})$, then there are three subcases, we give proof for only one subcase and the other two cases follow from this.\\
 If $i\notin \mathcal{A}_{1}$ and $i\in \mathcal{A}_{1}\Delta\mathcal{A}_{2}\Delta\mathcal{A}_{3}$, then
\begin{align*}
	&\widehat{g+h}({\bf 0})+\widehat{f}({\bf x})-\widehat{f+g+h}(x)\neq 2^{n}\Leftrightarrow 2^{n}-2\chi_{23}(2^{t}-1)+2s_{1}+2^{t+1}-2\chi_{123}\neq 2^{n}\\
		& \Leftrightarrow s_{12}+s_{13}-2s_{123}\neq 2^{t-1}(\chi_{23}-1)
\end{align*}
{\bf Case VI:} If $\phi_{1}=f$ and $\phi_{2}=f+g+h$, then for ${\bf x}$, we have following subcases:\\
(a) If ${\bf x}={\bf 0}$, then
\begin{align*}
	&\widehat{f}({\bf 0})+\widehat{f+g+h}({\bf 0})-\widehat{g+h}({\bf 0})\neq 2^{n}\Leftrightarrow 2^{n}-2s_{1}(2^{t}-1)+2^{n}-2\chi_{123}(2^{t}-1)-2^{n}+2\chi_{23}(2^{t}-1)\\
	& \Leftrightarrow -2s_{1}+2s_{12}+2s_{13}-4s_{123}\neq 0 \Leftrightarrow s_{1}\neq s_{12}+s_{13}-2s_{123}
\end{align*}
(b) If ${\bf x}\notin W_{i}^{\perp}$, for all $i\in (\mathcal{A}_{1}\Delta\mathcal{A}_{2}\Delta\mathcal{A}_{3})\cup (\mathcal{A}_{2}\Delta\mathcal{A}_{3})$, then
\begin{align*}
	&\widehat{f}({\bf 0})+\widehat{f+g+h}({\bf x})-\widehat{g+h}({\bf x})\neq 2^{n}\Leftrightarrow 2^{n}-2s_{1}(2^{t}-1)+2\chi_{123}-2\chi_{23}\neq 2^{n}\\
	& \Leftrightarrow
	s_{1}(1-2^{t-1})\neq (s_{12}+s_{13}-2s_{123})
\end{align*}
(c) If ${\bf x}\in W_{i}^{\perp}$ for some $i\in(\mathcal{A}_{1}\Delta\mathcal{A}_{2}\Delta\mathcal{A}_{3})\cup (\mathcal{A}_{2}\Delta\mathcal{A}_{3})$, then there are three subcases, we give proof for only one subcase and the other two cases follow from this.\\
If $i\notin \mathcal{A}_{1}\Delta\mathcal{A}_{2}\Delta\mathcal{A}_{3}$  and $i\in \mathcal{A}_{2}\Delta\mathcal{A}_{3}$, then
\begin{align*}
	&\widehat{f}({\bf 0})+\widehat{f+g+h}({\bf x})-\widehat{g+h}({\bf x})\neq 2^{n}\Leftrightarrow 2^{n}-2s_{1}(2^{t}-1)+2\chi_{123}+2^{t+1}-2\chi_{23}\neq 2^{n}\\
	& \Leftrightarrow s_{1}-(s_{12}+s_{13}-2s_{123})\neq 2^{t-1}(s_{1}-1)
\end{align*}
{\bf Case VII:} If $\phi_{1}=f+g+h$ and $\phi_{2}=f$, then for ${\bf x}$, we have following subcases:\\
(a) If ${\bf x}={\bf 0}$, it is follows from above.\\
(b) If ${\bf x}\notin W_{i}^{\perp}$, for all $i\in \mathcal{A}_{1}\cup(\mathcal{A}_{2}\Delta\mathcal{A}_{3})$, then
\begin{align*}
	&\widehat{f+g+h}({\bf 0})+\widehat{f}({\bf x})-\widehat{g+h}({\bf x})\neq 2^{n}\Leftrightarrow 2^{n}-2\chi_{123}(2^{t}-1)+2s_{1}-2\chi_{23}\neq 2^{n}\\
	& \Leftrightarrow
	 s_{1}-(s_{12}+s_{13}-2s_{123})\neq 2^{t-1}\chi_{123}
\end{align*}
(c) If ${\bf x}\in W_{i}^{\perp}$ for some $i\in\mathcal{A}_{1}\cup(\mathcal{A}_{2}\Delta\mathcal{A}_{3})$, then there are three subcases, we give proof for only one subcase and the other two cases follow from this.\\
If $i\notin \mathcal{A}_{1}$ and $i\in \mathcal{A}_{2}\Delta\mathcal{A}_{3}$, then
\begin{align*}
	&\widehat{f+g+h}({\bf 0})+\widehat{f}({\bf x})-\widehat{g+h}(x)\neq 2^{n}\Leftrightarrow 2^{n}-2\chi_{123}(2^{t}-1)+2s_{1}+2^{t+1}-2\chi_{23}\neq 2^{n}\\
	&\Leftrightarrow s_{1}-(s_{12}+s_{13}-2s_{123})\neq  2^{t-1}(\chi_{123}-1)
\end{align*}
{\bf Case VIII:} If $\phi_{1}=f+g$ and $\phi_{2}=f+g+h$, then for ${\bf x}$, we have following subcases:\\
(a) If ${\bf x}={\bf 0}$
\begin{align*}
	&\widehat{f+g}({\bf 0})+\widehat{f+g+h}({\bf 0})-\widehat{h}({\bf 0})\neq 2^{n}\\
	&\Leftrightarrow 2^{n}-2\chi_{12}(2^{t}-1)+2^{n}-2\chi_{123}(2^{t}-1)-2^{n}+2s_{3}(2^{t}-1)\neq 2^{n} \Leftrightarrow
	(s_{13}+s_{23}-2s_{123})\neq \chi_{12}
\end{align*}
(b) If ${\bf x}\notin W_{i}^{\perp}$, for all $i\in (\mathcal{A}_{1}\Delta\mathcal{A}_{2}\Delta\mathcal{A}_{3})\cup \mathcal{A}_{3}$, then
\begin{align*}
	&\widehat{f+g}({\bf 0})+\widehat{f+g+h}({\bf x})-\widehat{h}({\bf x})\neq 2^{n}\\
	&\Leftrightarrow 2^{n}-2\chi_{12}(2^{t}-1)+2\chi_{123}-2s_{3}\neq 2^{n}\Leftrightarrow
	 \chi_{12}-(s_{13}+s_{23}-2s_{123})\neq 2^{t-1}\chi_{12}
\end{align*}
(c) If ${\bf x}\in W_{i}^{\perp}$ for some $i\in(\mathcal{A}_{1}\Delta\mathcal{A}_{2}\Delta\mathcal{A}_{3})\cup \mathcal{A}_{3}$, then there are three subcases, we give proof for only one subcase and the other two cases follow from this.\\
If $i\notin \mathcal{A}_{1}\Delta\mathcal{A}_{2}\Delta\mathcal{A}_{3}$ and $i\in \mathcal{A}_{3}$, then
\begin{align*}
	&\widehat{f+g}({\bf 0})+\widehat{f+g+h}({\bf x})-\widehat{h}({\bf x})\neq 2^{n}\\
	&\Leftrightarrow 2^{n}-2\chi_{12}(2^{t}-1)+2\chi_{123}+2^{t+1}-2s_{3}\neq 2^{n}
	\Leftrightarrow 2^{t-1}(\chi_{12}-1)\neq\chi_{12}-(s_{13}+s_{23}-2s_{123})
\end{align*}
{\bf Case IX:} If $\phi_{1}=f+g+h$ and $\phi_{2}=f+g$, then for ${\bf x}$, we have following subcases:\\
(a) If ${\bf x}={\bf 0}$, then this case follows from above.\\
(b) If ${\bf x}\notin W_{i}^{\perp}$, for all $i\in (\mathcal{A}_{1}\Delta\mathcal{A}_{2})\cup\mathcal{A}_{3}$, then
\begin{align*}
	&\widehat{f+g+h}({\bf 0})+\widehat{f+g}({\bf x})-\widehat{h}({\bf x})\neq 2^{n}\\
	&\Leftrightarrow 2^{n}-2\chi_{123}(2^{t}-1)+2\chi_{12}-2s_{3}\neq 2^{n}\Leftrightarrow
	\chi_{123}(2^{t}-1)+s_{3}\neq \chi_{12}\\
	&[\text{Since } \max{(\chi_{12})}=2^{t}-2 \text{ and } \min{(\chi_{123}(2^{t}-1)+s_{3})}=2^{t+1}-2+2 ]
\end{align*}
(c) If ${\bf x}\in W_{i}^{\perp}$, for some $i\in(\mathcal{A}_{1}\Delta\mathcal{A}_{2})\cup\mathcal{A}_{3}$, then there are three subcases, we give proof for only one subcase and the other two cases follow from this.\\
If $i\notin \mathcal{A}_{1}\Delta\mathcal{A}_{2}$ and $i\in \mathcal{A}_{3}$,  then
\begin{align*}
	&\widehat{f+g+h}({\bf 0})+\widehat{f+g}({\bf x})-\widehat{h}({\bf x})\neq 2^{n}\\
	&\Leftrightarrow 2^{n}-2\chi_{123}(2^{t}-1)+2\chi_{12}+2^{t+1}-2s_{3}\neq 2^{n}
	\Leftrightarrow \chi_{123}(2^{t}-1)+s_{3}\neq \chi_{12}+2^{t}\\
	&[\text{Since } \max{(\chi_{12})}+2^{t}=2^{t}-2+2^{t} \text{ and } \min{(\chi_{123}(2^{t}-1)+s_{3})}=2^{t+1}-2+2 ]
\end{align*}
All other cases follow in the similar manner.\\
 Hence, $\widehat{\phi}_{1}({\bf 0})+\widehat{\phi}_{2}({\bf x})-\widehat{(\phi_{1}+\phi_{2})}({\bf x})\neq 2^{n}$, for all $\phi_{1},\phi_{2}\in \mathcal{F}$, with $\phi_{1}\neq \phi_{2}$ and ${\bf x},{\bf y}\in \mathbb{F}_{2}^{n}$, for all ${\bf x}\in \mathbb{F}_{2}^{n}\setminus\{{\bf 0}\}$.
\end{proof}
\end{proposition}
\begin{proposition}\label{lemma43}
	Let $n\geq 6$ and $2\leq s\leq 2^{t-1}$, for $s\in \{s_{1},s_{2},s_{3}\}$ and $\mathcal{F}$ be a collection of Boolean functions defined in \ref{eq41} and \ref{eq42}. Then for any ${\bf x}\in \mathbb{F}_{2}^{n}\setminus\{{\bf 0}\}$, we have
	\begin{equation*}
		\widehat{\phi}_{1}({\bf x})+\widehat{\phi}_{2}({\bf x})-\widehat{\phi_{1}+\phi_{2}}({\bf 0})\neq 2^{n}
	\end{equation*}
	for all, $\phi_{1},\phi_{2}\in \mathcal{F}$ with $\phi_{1}\neq \phi_{2}$.
\begin{proof}
For different choices of $\phi_{1}$ and $\phi_{2}$, we have following cases:\\
{\bf Case I:} For $\phi_{1}=f$ and $\phi_{2}=g$. Now, according to choice of ${\bf x}$, we have following possibilities:\\
(a) If ${\bf x}\notin W_{i}$ for all $i\in \mathcal{A}_{1}\cup \mathcal{A}_{2}$, then
\begin{align*}
&\widehat{f}({\bf x})+\widehat{g}({\bf x})-\widehat{f+g}({\bf 0})\neq 2^{n}\Leftrightarrow 2s_{1}+2s_{2}-2^{n}+2\chi_{12}(2^{t}-1)\neq 2^{n}\Leftrightarrow s_{12}\neq -2^{t-1}(2^{t}-\chi_{12})
\end{align*}
(b) If ${\bf x}\in W_{i}^{\perp}$ for some $i\in\mathcal{A}_{1}\cup \mathcal{A}_{2}$, then we divide this case into three subcases\\
(i) If $i \in \mathcal{A}_{1}\cap \mathcal{A}_{2}$,  then
\begin{align*}
	&\widehat{f}({\bf x})+\widehat{g}({\bf x})-\widehat{f+g}({\bf 0})\neq 2^{n}\Leftrightarrow -2^{t+1}+2s_{1}-2^{t+1}+2s_{2}-2^{n}+2\chi_{12}(2^{t}-1)\neq 2^{n}\\
	&\Leftrightarrow  2^{t}\chi_{12}-2s_{12}\neq 2^{n}+2^{t+1}
	\Leftrightarrow 2^{t-1}\chi_{12}\neq 2^{2t-1}+2^{t}+s_{12}
\end{align*}
(ii) If $i \in \mathcal{A}_{1}/ \mathcal{A}_{2}$, then
\begin{align*}
	&\widehat{f}({\bf x})+\widehat{g}({\bf x})-\widehat{f+g}({\bf 0})\neq 2^{n}\Leftrightarrow -2^{t+1}+2s_{1}+2s_{2}-2^{n}+2\chi_{12}(2^{t}-1)\neq 2^{n}\\
	&\Leftrightarrow  2^{t}\chi_{12}-2s_{12}\neq 2^{n}
	\Leftrightarrow 2^{t-1}\chi_{12}\neq 2^{2t-1}+2^{t-1}+s_{12}
\end{align*}
(iii) If $i \in \mathcal{A}_{2}/ \mathcal{A}_{1}$, then  this is the same case as above.\\
{\bf Case II:} For $\phi_{1}=f$ and $\phi_{2}=f+g$. Now, according to choice of ${\bf x}$, we have following possibilities:\\
(a) If ${\bf x}\notin W_{i}$, for all $i\in \mathcal{A}_{1}\cup (\mathcal{A}_{1}\Delta \mathcal{A}_{2})$, then
\begin{align*}
	&\widehat{f}({\bf x})+\widehat{f+g}({\bf x})-\widehat{g}({\bf 0})\neq 2^{n}\Leftrightarrow 2s_{1}+2\chi_{12}-2^{n}+2s_{2}(2^{t}-1)\neq 2^{n}\Leftrightarrow   s_{1}-s_{12}\neq 2^{t-1}(2^{t}-s_{2})
\end{align*}
(b) If ${\bf x}\in W_{i}^{\perp}$, for some $i\in\mathcal{A}_{1}\cup (\mathcal{A}_{1}\Delta \mathcal{A}_{2})$, then we divide this case into three subcases\\
(i) If $i \in \mathcal{A}_{1}\cap(\mathcal{A}_{1}\Delta\mathcal{A}_{2})$, then
\begin{align*}
	&\widehat{f}({\bf x})+\widehat{f+g}({\bf x})-\widehat{g}({\bf 0})\neq 2^{n}\Leftrightarrow -2^{t+1}+2s_{1}-2^{t+1}+2\chi_{12}-2^{n}+2s_{2}(2^{t}-1)\neq 2^{n}\\
	&\Leftrightarrow s_{1}-s_{12}\neq 2^{t-1}(2^{t}-s_{2}+2)
\end{align*}
(ii) If $i \in \mathcal{A}_{1}$ and $i\notin \mathcal{A}_{1}\Delta\mathcal{A}_{2}$, then
\begin{align*}
	&\widehat{f}({\bf x})+\widehat{f+g}({\bf x})-\widehat{g}({\bf 0})\neq 2^{n}\Leftrightarrow -2^{t+1}+2s_{1}+2
	\chi_{12}-2^{n}+2s_{2}(2^{t}-1)\neq 2^{n}\\
	&\Leftrightarrow s_{1}-s_{12}\neq 2^{t-1}(2^{t}-s_{2}+1)
\end{align*}
(iii) If $i \notin \mathcal{A}_{1}$ and $i\in \mathcal{A}_{1}\Delta\mathcal{A}_{2}$, then  this is the same case as above.\\
{\bf Case III:} For $\phi_{1}=f$ and $\phi_{2}=g+h$. Now, according to choice of ${\bf x}$, we have following possibilities:\\
(a) If ${\bf x}\notin W_{i}$, for all $i\in \mathcal{A}_{1}\cup (\mathcal{A}_{2}\Delta \mathcal{A}_{3})$, then
\begin{align*}
	&\widehat{f}({\bf x})+\widehat{g+h}({\bf x})-\widehat{f+g+h}({\bf 0})\neq 2^{n}\Leftrightarrow 2s_{1}+2\chi_{23}-2^{n}+2\chi_{123}(2^{t}-1)\neq 2^{n}\\
	&\Leftrightarrow  2^{t-1}\chi_{123}+(s_{12}+s_{13}+2s_{123})\neq 2^{n-1}\Leftrightarrow (s_{12}+s_{13}+2s_{123}) \neq 2^{t-1}(2^{t}-\chi_{123})
\end{align*}
(b) If ${\bf x}\in W_{i}^{\perp}$, for some $i\in\mathcal{A}_{1}\cup( \mathcal{A}_{2}\Delta\mathcal{A}_{3})$, then we divide this case into three subcases\\
(i) If $i \in \mathcal{A}_{1}\cap(\mathcal{A}_{2}\Delta\mathcal{A}_{3})$, then
\begin{align*}
	&\widehat{f}({\bf x})+\widehat{g+h}({\bf x})-\widehat{f+g+h}({\bf 0})\neq 2^{n}\Leftrightarrow -2^{t+1}+2s_{1}-2^{t+1}+2\chi_{23}-2^{n}+2\chi_{123}(2^{t}-1)\neq 2^{n}\\
	&\Leftrightarrow  2^{t-1}\chi_{123}+(s_{12}+s_{13}-2s_{123})\neq 2^{n-1}+2^{t}
	\Leftrightarrow (s_{12}+s_{13}+2s_{123}) \neq 2^{t-1}(2^{t}-\chi_{123})+2^{t}
\end{align*}
(ii) If $i \in \mathcal{A}_{1}$ and $i\notin \mathcal{A}_{2}\Delta\mathcal{A}_{3}$, then
\begin{align*}
	&\widehat{f}({\bf x})+\widehat{g+h}({\bf x})-\widehat{h}({\bf 0})\neq 2^{n}\Leftrightarrow -2^{t+1}+2s_{1}+2
	\chi_{23}-2^{n}+2\chi_{123}(2^{t}-1)\neq 2^{n}\\
	&\Leftrightarrow 2^{t-1}\chi_{123}+(s_{12}+s_{13}+2s_{123})\neq 2^{n-1}+2^{t-1}
	\Leftrightarrow  (s_{12}+s_{13}+2s_{123}) \neq 2^{t-1}(2^{t}-\chi_{123})+2^{t-1}
\end{align*}
(iii) If $i \notin \mathcal{A}_{1}$ and $i\in \mathcal{A}_{2}\Delta\mathcal{A}_{3}$, then  this is the same case as above.\\
{\bf Case IV:} For $\phi_{1}=f$ and $\phi_{2}=f+g+h$. Now, according to choice of ${\bf x}$, we have following possibilities:\\
(a) If ${\bf x}\notin W_{i}$ for all $i\in \mathcal{A}_{1}\cup( \mathcal{A}_{1}\Delta\mathcal{A}_{2}\Delta \mathcal{A}_{3})$, then
\begin{align*}
	&\widehat{f}({\bf x})+\widehat{f+g+h}({\bf x})-\widehat{g+h}({\bf 0})\neq 2^{n}\Leftrightarrow 2s_{1}+2\chi_{123}-2^{n}+2\chi_{23}(2^{t}-1)\neq 2^{n}\\
	&\Leftrightarrow 2s_{1}-2s_{12}-2s_{13}+4s_{123} \neq 2^{n}-2^{t}\chi_{23}\Leftrightarrow s_{1}-(s_{12}+s_{13}-2s_{123})\neq 2^{t-1}(2^{t}-\chi_{23})
\end{align*}
(b) If ${\bf x}\in W_{i}^{\perp}$ for some $i\in\mathcal{A}_{1}\cup( \mathcal{A}_{1}\Delta\mathcal{A}_{2}\Delta \mathcal{A}_{3})$, then we divide this case into three subcases\\
(i) If $i \in \mathcal{A}_{1}\cap(\mathcal{A}_{1}\Delta\mathcal{A}_{2}\Delta\mathcal{A}_{3})$, then
\begin{align*}
	&\widehat{f}({\bf x})+\widehat{f+g+h}({\bf x})-\widehat{g+h}({\bf 0})\neq 2^{n}\Leftrightarrow -2^{t+1}+2s_{1}-2^{t+1}+2\chi_{123}-2^{n}+2\chi_{23}(2^{t}-1)\neq 2^{n}\\
	&\Leftrightarrow 2s_{1}-2s_{12}-2s_{13}+4s_{123} \neq 2^{n}-2^{t}\chi_{23}+2^{t}\Leftrightarrow s_{1}-(s_{12}+s_{13}-2s_{123})\neq 2^{t-1}(2^{t}-\chi_{23})+2^{t}
\end{align*}
(ii) If $i \in \mathcal{A}_{1}$ and $i\notin \mathcal{A}_{1}\Delta\mathcal{A}_{2}\Delta\mathcal{A}_{3}$, then
\begin{align*}
	&\widehat{f}({\bf x})+\widehat{f+g+h}({\bf x})-\widehat{g+h}({\bf 0})\neq 2^{n}\Leftrightarrow -2^{t+1}+2s_{1}+2\chi_{123}-2^{n}+2\chi_{23}(2^{t}-1)\neq 2^{n}\\
	&\Leftrightarrow 2s_{1}-2s_{12}-2s_{13}+4s_{123} \neq 2^{n}-2^{t}\chi_{23}+2^{t}\Leftrightarrow s_{1}-(s_{12}+s_{13}-2s_{123})\neq 2^{t-1}(2^{t}-\chi_{23})+2^{t-1}
\end{align*}
(iii) If $i \notin \mathcal{A}_{1}$ and $i\in\mathcal{A}_{1}\Delta \mathcal{A}_{2}\Delta\mathcal{A}_{3}$, then  this is the same case as above.\\
{\bf Case V:} For $\phi_{1}=f+g$ and $\phi_{2}=f+g+h$. Now, according to choice of ${\bf x}$, we have following possibilities:\\
(a) If ${\bf x}\notin W_{i}$ for all $i\in (\mathcal{A}_{1}\Delta\mathcal{A}_{2})\cup(\mathcal{A}_{1}\Delta \mathcal{A}_{2}\Delta \mathcal{A}_{3})$, then
\begin{align*}
	&\widehat{f+g}({\bf x})+\widehat{f+g+h}({\bf x})-\widehat{h}({\bf 0})\neq 2^{n}\Leftrightarrow 2\chi_{12}+2\chi_{123}-2^{n}+2s_{3}(2^{t}-1)\neq 2^{n}\\
	&\Leftrightarrow 2\chi_{12}-2(s_{13}+s_{23}-2s_{123}) \neq 2^{n}-2^{t}s_{3}\Leftrightarrow \chi_{12}-(s_{13}+s_{23}-2s_{123}) \neq 2^{t-1}(2^{t}-s_{3})
\end{align*}
(b) If ${\bf x}\in W_{i}^{\perp}$ for some $i\in(\mathcal{A}_{1}\Delta\mathcal{A}_{2})\cup(\mathcal{A}_{1}\Delta \mathcal{A}_{2}\Delta \mathcal{A}_{3})$, then we divide this case into three subcases\\
(i) If $i \in (\mathcal{A}_{1}\Delta\mathcal{A}_{2})\cap (\mathcal{A}_{1}\Delta\mathcal{A}_{2}\Delta\mathcal{A}_{3})$, then
\begin{align*}
	&\widehat{f+g}({\bf x})+\widehat{f+g+h}({\bf x})-\widehat{h}({\bf 0})\neq 2^{n}\Leftrightarrow -2^{t+1}+2\chi_{12}-2^{t+1}+2\chi_{123}-2^{n}+2s_{3}(2^{t}-1)\neq 2^{n}\\
&\Leftrightarrow \chi_{12}-(s_{13}+s_{23}-2s_{123}) \neq 2^{t-1}(2^{t}-s_{3})+2^{t}
\end{align*}
(ii) If $i \in \mathcal{A}_{1}\Delta\mathcal{A}_{2}$ and $i\notin \mathcal{A}_{1}\Delta\mathcal{A}_{2}\Delta\mathcal{A}_{3}$, then
\begin{align*}
	&\widehat{f+g}({\bf x})+\widehat{f+g+h}({\bf x})-\widehat{h}({\bf 0})\neq 2^{n}\Leftrightarrow -2^{t+1}+2\chi_{12}-2^{t+1}+2\chi_{123}-2^{n}+2s_{3}(2^{t}-1)\neq 2^{n}\\
	&\Leftrightarrow \chi_{12}-(s_{13}+s_{23}-2s_{123}) \neq 2^{t-1}(2^{t}-s_{3})+2^{t-1}
\end{align*}
(iii) If $i \notin \mathcal{A}_{1}\Delta\mathcal{A}_{2}$ and $i\in\mathcal{A}_{1}\Delta \mathcal{A}_{2}\Delta\mathcal{A}_{3}$, then  this is the same case as above.
All other cases follow in the similar manner.\\ Henec, $\widehat{\phi}_{1}({\bf x})+\widehat{\phi}_{2}({\bf x})-\widehat{\phi_{1}+\phi_{2}}({\bf 0})\neq 2^{n}$, for all $\phi_{1},\phi_{2}\in \mathcal{F}$ with $\phi_{1}\neq\phi_{2} $ and ${\bf x}\in \mathbb{F}_{2}^{n}\setminus\{{\bf 0}\}$.
\end{proof}
\end{proposition}

\begin{proposition}\label{lemma45}
	Let $n\geq 6$ and $2\leq s\leq 2^{t-1}$, for $s\in \{s_{1},s_{2},s_{3}\}$ and $\mathcal{F}$ be an collection of Boolean functions defined in \ref{eq41} and \ref{eq42}. Then for any ${\bf x},{\bf y}\in \mathbb{F}_{2}^{n}\setminus\{{\bf 0}\}$ with ${\bf x}\neq {\bf y}$, we have
	\begin{equation*}
		\widehat{\phi}_{1}({\bf x}+{\bf y})+\widehat{\phi}_{2}({\bf x})-\widehat{\phi_{1}+\phi_{2}}({\bf y})\neq 2^{n}
	\end{equation*}
	for all, $\phi_{1},\phi_{2}\in \mathcal{F}$ with $\phi_{1}\neq \phi_{2}$.
\begin{proof}
Denote $\psi_{1}=\widehat{\phi}_{1}({\bf x}+{\bf y})+\widehat{\phi}_{2}({\bf x})-\widehat{\phi_{1}+\phi_{2}}({\bf y})$. As, ${\bf x}$ and ${\bf y}$ are non-zero vectors with ${\bf x}\neq {\bf y}$, this implies ${\bf x}+{\bf y}\neq 0$. Now, for different choice of $\phi_{1}$ and $\phi_{2}$, we have following cases:\\
{\bf Case I:} If $\phi_{1}=f$ and $\phi_{2}=g$, then	 $\psi_{1}\in \{s-2^{t+2},s-2^{t+1},s, s+2^{t+1}\}$, where $s=4s_{12}$. We know that, $s_{12}<2^{t-1}$, from this, we get $\psi_{1}<2^{n}$. \\
{\bf Case II:} If $\phi_{1}=f$ and $\phi_{2}=f+g$, then  $\psi_{1}\in\{s-2^{t+2},s-2^{t-1},s, s+2^{t+1}\}$, where $s=4s_{1}-4s_{12}$. We know that, $s_{1}\leq 2^{t-1}$, from this, we get $\psi_{1}<2^{n}$.\\
{\bf Case III:} If $\phi_{1}=f$ and $\phi_{2}=g+h$, then	 $\psi_{1}\in \{s-2^{t+2},s-2^{t+1},s, s+2^{t+1}\}$, where $s=4(s_{12}+s_{13}-2s_{123})$. We know that, $(s_{12}+s_{13}-2s_{123})<2^{t-1}$, from this, we get $\psi_{1}<2^{n}$. \\
{\bf Case IV:} If $\phi_{1}=f$ and $\phi_{2}=f+g+h$, then	 $\psi_{1}\in \{s-2^{t+2},s-2^{t+1},s, s+2^{t+1}\}$, where $s=4(s_{1}-(s_{12}+s_{13}-2s_{123}))$. We know that, $(s_{1}-(s_{12}+s_{13}-2s_{123}))<2^{t-1}$, from this, we get $\psi_{1}<2^{n}$. \\
{\bf Case V:} If $\phi_{1}=f+g$ and $\phi_{2}=f+g+h$, then	 $\psi_{1}\in \{s-2^{t+2},s-2^{t+1},s, s+2^{t+1}\}$, where $s=4((s_{1}+s_{2}-2s_{12})-(s_{13}+s_{23}-2s_{123}))$. We know that, $(s_{1}+s_{2}-2s_{12})\leq2^{t}-2$, from this, we get $\psi_{1}<2^{n}$. \\
All other cases follow in the similar manner.\\
 Hence, $\widehat{\phi}_{1}({\bf x}+{\bf y})+\widehat{\phi}_{2}({\bf x})-\widehat{\phi_{1}+\phi_{2}}({\bf y})\neq 2^{n}$, for all $\phi_{1},\phi_{2}\in \mathcal{F}$ with $\phi_{1}\neq \phi_{2}$ and ${\bf x},{\bf y}\in \mathbb{F}_{2}^{n}\setminus\{0\}$, with ${\bf x}\neq {\bf y}$.
\end{proof}
\end{proposition}
The following result gives a family of minimal linear codes of dimension $n+3$ violating the Ashikhmin-Barg condition.
\begin{theorem}
Let $n\geq 6$ and $2\leq s\leq 2^{t-1}$, for $s\in \{s_{1},s_{2},s_{3}\}$. If $\epsilon\leq 2^{t-1}-1$, where $\epsilon=\min{\{s_{1},s_{2},s_{3},\chi_{12},\chi_{13},\chi_{23},\chi_{123}\}}$, then the code $\mathscr{C}_{f,g,h}$ is $[2^{n-1},n+3,\epsilon(2^{t}-1)]$ a minimal linear code. Moreover, if $\epsilon\leq 2^{t-2}$, then $\frac{{wt}_{min}}{{wt}_{max}}\leq \frac{1}{2}$.
\begin{proof}
We first prove that $\mathscr{C}_{f,g,h}$ is minimal linear code by Theorem \ref{thm32}. From Proposition \ref{lemma41} and \ref{lemma46}, Theorem \ref{thm32}(1) is true for $\mathscr{C}_{f,g,h}$. For Theorem \ref{eq31}(2), we divide proof into three cases as follows:\\
{\bf Case I:} If ${\bf x}={\bf y}\in \mathbb{F}_{2}^{n}$, then by Proposition \ref{lemma42}, Theorem \ref{eq31}(2) is true for $\mathscr{C}_{f,g,h}$.\\
{\bf Case II:} If one of ${\bf x}$ and ${\bf y}$ is zero,  then from Proposition \ref{lemma42} and \ref{lemma43}, we get Theorem \ref{eq31}(2) holds.\\
{\bf Case III:} If ${\bf x}$ and ${\bf y}$ are different non-zero vectors, then Theorem \ref{eq31}(2) holds   by Proposition \ref{lemma45}.\\
Hence, $\mathscr{C}_{f,g,h}$ is a minimal linear code. Note that, possible values for minimum weight is $\epsilon(2^{t}-1)$ and $2^{m-1}-\mu$, where $\mu=\max{\{s_{1},s_{2},s_{3},\chi_{12},\chi_{13},\chi_{23},\chi_{123}\}}$. We know that, possibly large value of $\mu$ is $=2^{t}-2$ and  $\epsilon(2^{t}-1)\leq 2^{m-1}-\mu$. From this, we get ${wt}_{min}=\epsilon(2^{t}-1)$ and $wt_{max}=2^{m-1}+2^{t}-\epsilon(2^{t}-1)$. We can easily prove that $\frac{wt_{min}}{wt_{max}}\leq \frac{1}{2}$.\\
This conclude the result.
\end{proof}
\end{theorem}
\section{Conclusion}
In \cite{13} Liu, H., \& Liao, Q. constructed a minimal binary linear code violating the Ashikhmin-Barg condition of dimension $n+2$ and using the procedure from this paper, we extend and generalize this construction of a minimal binary linear code of dimension $n+3$. We can think of this as an algorithm to increase the dimension of minimal linear code.
In this paper, we gave a generic construction of a binary linear code of dimension $n+3$ and stated the necessary and sufficient condition for the constructed code to be minimal. Using this condition, we constructed a class of minimal binary linear code from a class of special Boolean functions violating the Ashikhmin-Barg condition. Also, we obtained the weight distribution of the constructed minimal binary linear code.
\par  Finding a minimal linear code of dimension greater than $n+3$ and extending the construction in \ref{eq1.1} for non-binary fields to construct a minimal linear code violating the Ashikhmin-Barg condition is the further scope of the study.

\end{document}